\documentclass[aps, pra, amsfonts, amssymb, amsmath, twocolumn]{revtex4-1}

\pdfoutput=1

\usepackage{amsfonts}
\usepackage{amssymb}
\usepackage{amsmath}
\usepackage{hyperref}           % Highlights the references blue.
\hypersetup{colorlinks = True, 
            allcolors = {blue}
}
\usepackage{natbib}
\usepackage{relsize}            % for resizing letters in equations
\usepackage{qcircuit}           % quantum circuits
\usepackage{graphicx}                  % figures 
\usepackage[caption=false]{subfig}     % figures 

\newcommand{\bra}[1]{ \langle{#1}| }
\newcommand{\ket}[1]{ |{#1}\rangle }

\def\rVac{ |\text{vac} \rangle }
\def\rsVac{ | \bar{0} \rangle }

\def\rAGP{|\text{AGP}\rangle}
\def\lAGP{\langle \text{AGP} |}
\def\AGPnorm{\langle \text{AGP}|\text{AGP} \rangle}
\def\rBCS{|\text{BCS}\rangle}
\def\lBCS{\langle \text{BCS} |}
\def\BCSnorm{\langle \text{BCS}|\text{BCS} \rangle}

\def\rOn{|1\rangle}
\def\rOff{|0\rangle}

\newcommand{\Cdag}[1]{ {c}^{\dagger}_{#1} }
\newcommand{\Cp}[1]{ {c}_{#1} }

\def\GamD{{\mathbf{{\Gamma}}}^{\dagger}}
\newcommand{\Pdag}[1]{\mathbf{P}^{\dagger}_{#1} }
\newcommand{\N}[1]{ \mathbf{N}_{#1} }
\newcommand{\Pp}[1]{ \mathbf{P}_{#1} }
\newcommand{\Killer}[1]{\mathbf{K}_{#1}}
\newcommand{\KillerAd}[1]{\mathbf{K}^{\dagger}_{#1}}
\newcommand{\PHoppers}{\hat{\mathcal{T}}}

\newcommand{\Sx}[1]{ {X}_{#1} }
\newcommand{\Sy}[1]{ {Y}_{#1} }
\newcommand{\Sz}[1]{ {Z}_{#1} }
\newcommand{\Su}[1]{ \sigma^{+}_{#1} }
\newcommand{\Sd}[1]{ \sigma^{-}_{#1} }

\newcommand{\E}[1]{\langle {#1} \rangle}
\newcommand{\Eagp}[1]{\lAGP {#1} \rAGP}

\newcommand{\Proj}{\mathcal{P}_N}
\newcommand{\ProjOp}[1]{\hat{\mathcal{R}}(#1)}

\newcommand{\icol}[1]{\left(\begin{smallmatrix}#1\end{smallmatrix}\right)}

\newcommand{\bigO}[1]{$\mathcal{O}(#1)$}
\newcommand{\floor}[1]{\lfloor #1 \rfloor}
\newcommand{\Ham}{\mathlarger{\hat{H}}}

\newcommand{\Eq}[1]{Eq.~({#1})}
\newcommand{\Sec}[1]{Sec.~{#1}}
\newcommand{\Fig}[1]{Fig.~{#1}}
\newcommand{\Reference}[1]{Ref.~{#1}}

\begin{document}

\title{Correlating AGP on a quantum computer}
\author{Armin Khamoshi}
    \email[Correspondence email address: ]{armin.khamoshi@rice.edu}
    \affiliation{Department of Physics and Astronomy, Rice University, Houston, TX 77005-1892}
    
\author{Francesco A. Evangelista}
    \affiliation{Department of Chemistry and Cherry Emerson Center for Scientific Computation, Emory University, Atlanta, GA, 30322}
    
\author{Gustavo E. Scuseria}
    \affiliation{Department of Chemistry, Rice University, Houston, TX 77005-1892}
    \affiliation{Department of Physics and Astronomy, Rice University, Houston, TX 77005-1892}

\date{\today} 

\begin{abstract}
For variational algorithms on the near term quantum computing hardware, it is highly desirable to use very accurate ansatze with low implementation cost. Recent studies have shown that the antisymmetrized geminal power (AGP) wavefunction can be an excellent starting point for ansatze describing systems with strong pairing correlations, as those occurring in superconductors. In this work, we show how AGP can be efficiently implemented on a quantum computer with circuit depth, number of CNOTs, and number of measurements being linear in system size. Using AGP as the initial reference, we propose and implement a unitary correlator on AGP and benchmark it on the ground state of the pairing Hamiltonian. The results show highly accurate ground state energies in all correlation regimes of this model Hamiltonian. 
\end{abstract} 


\maketitle

%====================================================================%
%                      ** INTRODUCTION **                            %
%====================================================================%
\section{Introduction}
Recent advances in quantum computing have opened up new avenues to tackle the strong correlation problem in electronic structure theory. \cite{babbush_low-depth_2018, cao_quantum_2019, mcardle_quantum_2020} Due to the exponential cost of simulating the entire Hilbert space on a digital computer, a quantum computer with as few as 50--100 qubits could in principle outperform their classical counterparts in some tasks. \cite{preskill_quantum_2018, arute_quantum_2019, mcardle_quantum_2020} However, existing noisy intermediate scale quantum (NISQ) devices, suffer from short coherence-time and lack of error correction. \cite{li_efficient_2017, temme_error_2017, preskill_quantum_2018, mcardle_quantum_2020} Therefore, while there exist algorithms such as quantum phase estimation (QPE) that can compute the ground state energies of any fermionic system with exponential speedup, \cite{abrams_simulation_1997, abrams_quantum_1999} they cannot be reliably implemented on NISQ devices. A promising alternative is to use hybrid quantum-classical algorithms, chief among which is the variational quantum eigensolver (VQE) \cite{peruzzo_variational_2014, mcclean_theory_2016}. In VQE, the ground state of a Hamiltonian, $\Ham$, is obtained by variationally optimizing the energy over an ansatz $\ket{\psi(\theta)}$ that depends on a set of parameters $\theta$. That is,  
\begin{align}
    E_{\text{gs}} = \underset{ \theta }{\text{min}} \; \bra{\psi(\theta)} \hat{H} \ket{\psi(\theta)},
\end{align}
such that the state preparation is done on a quantum computer, while the parameter optimization is performed on a classical computer. The role of a quantum computer in VQE is to overcome the exponential cost of storing the wavefunction, which would be intractable on a classical computer.

Choosing an appropriate ansatz in VQE is absolutely crucial in converging to or near the ground state. \cite{dallaire-demers_low-depth_2019, cao_quantum_2019, mcardle_quantum_2020} On the one hand, we demand that the physical resources needed for implementing $\ket{\psi(\theta)}$ should scale polynomially in system size and accuracy. On the other hand, we want the ansatz to have a large overlap with the ground state and guarantee that it can access the relevant parts of the Hilbert space  in the optimization. \cite{cao_quantum_2019, mcardle_quantum_2020, barron_preserving_2020, gard_efficient_2020} A common approach is to use variants of unitary coupled cluster singles and doubles (UCCSD) on the Hartree-Fock (HF) reference. \cite{bartlett_alternative_1989, kutzelnigg_error_1991, taube_new_2006, cooper_benchmark_2010, evangelista_alternative_2011, mcclean_theory_2016, barkoutsos_quantum_2018, romero_strategies_2018, harsha_difference_2018, grimsley_adaptive_2019, lee_generalized_2019} Such physically-inspired ansatze are typically more accurate than their ad-hoc, hardware-efficient counterparts, but they often require relatively deeper and more expensive circuits to implement. \cite{kandala_hardware-efficient_2017, mcclean_barren_2018, mcardle_quantum_2020, grimsley_is_2020, tang_qubit-adapt-vqe_2020} The cost is exacerbated in the strong correlation regime where collective excitations become important, which in turn requires even deeper circuits to implement. \cite{grimsley_adaptive_2019, lee_generalized_2019} Moreover, considerations concerning the so-called ``symmetry dilemma" could further complicate the applications of unitary coupled cluster in the presence of strong correlation. \cite{jimenez-hoyos_projected_2012, bulik_can_2015, tsuchimochi_exact_2020, lacroix_symmetry_2020}

Yet, there exist strongly correlated systems for which neither HF-based nor multireference methods might be the best starting points. Consider for example the attractive pairing---also known as the reduced Bardeen-Cooper-Schrieffer (BCS)---Hamiltonian \cite{bardeen_theory_1957, bayman_derivation_1960, sierra_exact_2000, dukelsky_colloquium:_2004}, which can be written as 
\begin{align} \label{eq:pairing_Ham_fermions}
    \hat{H} = \sum_p \epsilon_p (\hat{n}_{p\uparrow} + \hat{n}_{p\downarrow})
    - G \sum_{pq} \Cdag{p\uparrow}\Cdag{p\downarrow} \Cp{q\downarrow}\Cp{q\uparrow},
\end{align}
where $\Cdag{p\sigma}$ and $\hat{n}_{p\sigma}$ are the creation and number operators respectively of a fermion in orbital $p$ and spin $\sigma=\{\uparrow, \downarrow\}$. Here, and in the rest of the paper, we assume for simplicity that $\epsilon_p = p \Delta \epsilon$ is the single-particle energy level such that $\Delta\epsilon$ is the level-spacing, and $G$ is a constant that tunes the strength of the pairwise interaction. Note that the interaction is infinite-range, and it is attractive when $G>0$. The relevant symmetries of this Hamiltonian are seniority \cite{bytautas_seniority_2011} (i.e. each orbital is either doubly occupied or empty) and the total particle number. The lowest energy mean-field solution spontaneously breaks number symmetry in finite systems at some critical value $G = G_c>0$. This gives rise to the well-known BCS wavefunction \cite{bardeen_theory_1957} for $G>G_c$ and a symmetry preserving Slater determinant for all $G<G_c$. While this Hamiltonian is exactly solvable by the Richardson-Gaudin equations, \cite{richardson_restricted_1963, dukelsky_colloquium:_2004} some of the widely used many-body methods, e.g. coupled cluster theory, break down in the regime where the attractive interaction is strong. \cite{henderson_quasiparticle_2014, henderson_pair_2015, degroote_polynomial_2016, qiu_particle-number_2019, henderson_correlating_2020} In particular, it has been shown that neither symmetry-adapted nor broken-symmetry single-reference coupled cluster theory is a suitable approach to solve this problem. \cite{henderson_correlating_2020} Thus, by extension, one might conjecture their unitary counterparts are similarly ineffective. Multireference methods are not suitable either, because in the limit where $G \gg G_c$, all Slater determinants become equally important, which makes it impossible to select an active orbital space. \cite{henderson_correlating_2020} 

Meanwhile, the AGP wavefunction has emerged as an excellent starting point for this problem. \cite{henderson_geminal-based_2019,henderson_correlating_2020, dutta_geminal_2020} AGP, which is equivalent to the number-projected BCS wavefunction, \cite{ring_nuclear_1980, blaizot_quantum_1986, dukelsky_structure_2016} is well known for its ability to describe off-diagonal long-range order without breaking number symmetry. \cite{yang_concept_1962} While AGP is not necessarily a good wavefunction \textit{per se}, since geminals are not all the same in most physical problems, it has been shown recently that correlated wavefunctions built from AGP are good at describing both the weak and strong pairing correlations---at least in the reduced BCS Hamiltonian. \cite{henderson_geminal-based_2019, henderson_correlating_2020, dutta_geminal_2020} There are many qualities that could make AGP an attractive starting point for a more generic Hamiltonian wherein pairing correlations play a role. First, it inherently contains the same number of Slater determinants as doubly occupied configuration interaction (DOCI), \cite{veillard_complete_1967, couty_generalized_1997, kollmar_new_2003, bytautas_seniority_2011} yet it can be optimized with mean-field cost, i.e. \bigO{M^3} where $M$ is the system size. \cite{sheikh_symmetry-projected_2000, scuseria_projected_2011} (Note that DOCI is exact for Hamiltonians where seniority is a good quantum number, but it has combinatorial cost.) Moreover, AGP contains HF, thus it has a much richer structure as an initial reference. Secondly, many-body reduced density matrices (RDMs) can be computed efficiently over AGP. In particular, any $n$-body density matrix can be written as a linear combination of lower rank density matrices and geminal coefficients. \cite{khamoshi_efficient_2019}  Indeed, this is reminiscent of HF theory where all RDMs can be obtained from lower order ones.

In this paper, we propose an efficient algorithm to implement AGP on a quantum computer. That is, having decided to explore the use of AGP as an initial reference, we first optimize AGP on a classical computer, and then improve it on a quantum computer by a unitary correlator acting on it. Our method paves the way for taking advantage of unitary ansatze built atop of AGP, which are only accessible in approximate form on a classical computer. \cite{khamoshi_manuscript_nodate} In \Sec{\ref{sub:pair-hoppers}} we make use of an operator which we colloquially call unitary \textit{pair-hopper} and benchmark the ansatz by optimizing it for the pairing Hamiltonian. As we shall see in \Sec{\ref{sec:application}}, this ansatz is accurate not only for attractive interactions where the conventional methods break down, but it is also well-behaved for repulsive interactions where coupled cluster is accurate. While this work focuses on the pairing Hamiltonian, a seniority conserving model that we used as an initial step, future work aims to extend the present ideas to more general systems e.g. ab initio Hamiltonians.

%====================================================================%
%                         ** THEORY **                               %
%====================================================================%
\section{Theory} \label{sec:thoery}
Our strategy for implementing AGP on a quantum computer is to first efficiently simulate the corresponding BCS wavefunction and then number-project it in a NISQ-friendly manner. In \Sec{\ref{sub:geminals_to_Q}}, we take advantage of an economic mapping between fermion pairs and qubits which proves to be highly advantageous in reducing the number of qubits and making the circuits shallow. In \Sec{\ref{sub:BCS}}, we show an efficient implementation of the BCS wavefunction using single-qubit rotations only. Other   authors have discussed implementing a generic fermionic Gaussian state on a quantum computer \cite{jiang_quantum_2018, dallaire-demers_low-depth_2019}. Our approach differs from those methods in that we do not rely on a quasi-particle encoding and the Bogoliubov transformation. In \Sec{\ref{sub:NP_AGP}} we discuss a procedure to carry out number projection with the aid of a series of measurements. Finally, in \Sec{\ref{sub:pair-hoppers}} we derive our correlator from the killers of AGP and discuss its implementation.

%---------------------------------------------------------------------
\subsection{Mapping geminals to qubits} \label{sub:geminals_to_Q}
A geminal creation operator can be expressed as
\begin{align}
    \GamD = \sum_{p,q = 1}^{2M} \eta_{pq} \Cdag{p} \Cdag{q},
\end{align}
where $\Cdag{p}$ is the creation operator of a fermion in spin-orbital $p$, $\eta _{pq}$ is the geminal coefficient (an antisymmetric matrix), and there are a total of $2M$ spin-orbitals in the system. AGP with $N$ pairs is a geminal-based wavefunction where all $2N$ fermions are in the same geminal \cite{coleman_structure_1965}
\begin{align} \label{eq:AGP_as_geminal}
    \rAGP = \frac{1}{N!} (\GamD)^N \rVac,
\end{align}
where $\rVac$ is the physical vacuum and the $1/N!$ factor is introduced for convenience. 

While it is possible to implement the geminal operator on a quantum computer by mapping fermions to qubits using the Jordan-Wigner, Bravyi-Kitaev, or other transformations, \cite{jordan_uber_1928, bravyi_fermionic_2002, seeley_bravyi-kitaev_2012} we show that a more efficient implementation can be obtained by mapping each pair of fermions to a qubit. To this end, without loss of generality, we apply an orbital rotation that brings the matrix of the geminal coefficients into a block-diagonal form. \cite{hua_theory_1944} This expresses the geminal operator in the natural-orbital basis of the geminal wherein all orbitals are paired. Therefore, we can write the geminal operator readily in terms of pair creation operators
\begin{align}
    \GamD=\sum_{p=1}^{M} \eta _{p} \Pdag{p},
\end{align}
where we define
\begin{subequations} \label{eqs:hardcore_bosons}
\begin{align}
    \Pdag{p} &= \Cdag{p}\Cdag{\bar{p}}, \\
    \N{p}    &= \Cdag{p}\Cp{p} + \Cdag{\bar{p}}\Cp{\bar{p}},
\end{align}
\end{subequations}
such that $\bar{p}$ is the ``paired" companion of $p$. Therefore, the AGP wavefunction in \Eq{\ref{eq:AGP_as_geminal}} can be written as 
\begin{align}
    \rAGP = \sum_{ 1 \leq p_1<...<p_N \leq M} \eta _{p_1}...\eta _{p_N} \Pdag{p_1}...\Pdag{p_{N}} \rVac.
\end{align}

The operators, $\Pdag{p}$, $\N{p}$, and $\Pp{p}$ are generators of a $su(2)$ Lie algebra \cite{richardson_exact_1964, dukelsky_structure_2016, khamoshi_efficient_2019}
\begin{subequations}\label{eq:CommutationRelations}
\begin{align}
\left[\Pp{p}, \Pdag{q}\right] &= \delta _{pq}\left( 1- \N{p}\right), \\
\left[\N{p}, \Pp{q}^{\dagger}\right] &= 2\delta _{pq} \Pdag{q},
\end{align}
\end{subequations}
thus can be naturally mapped to the standard model of quantum computation \cite{ortiz_quantum_2001} as follows: Let $ \rOff_p = \icol{1\\0}$ and $\rOn_p = \icol{0\\1}$ represent the doubly-unoccupied and doubly-occupied natural orbital $p$ respectively, then
\begin{subequations} \label{eqs:pair-qubit-mapping}
\begin{align}
    \rVac    &\mapsto \rOff_M\otimes...\otimes\rOff_{1} \equiv \rsVac, \\
    \Pdag{p} &\mapsto \frac{1}{2} (\Sx{p} - i\Sy{p}) \equiv \Su{p},\\
    \N{p}    &\mapsto 1 - \Sz{p} \equiv n_p,\\
    \Pp{p}   &\mapsto \frac{1}{2} (\Sx{p} + i\Sy{p})\equiv \Sd{p},
\end{align}
\end{subequations}
where $X$, $Y$, and $Z$ are the standard Pauli operators. Note that the tensor products are ordered in such a way that those with smaller indices are placed on the right. 

It is easy to show that the two algebras are isomorphic. The advantage of this mapping is that we need half as many qubits ($M$ as opposed to the original $2M$) in the implementation. Moreover, since the pair operators commute for off-site indices, the Pauli Z strings associated with the anticommutation of fermions are absent. \cite{[{After submitting this manuscript, it has come to our attention that the following preprint used a slight variation of this mapping: }]elfving_simulating_2020}

%---------------------------------------------------------------------
\subsection{BCS wavefunction on a quantum computer} \label{sub:BCS}
Recall that the normalized BCS wavefunction can be written as \cite{bardeen_theory_1957}
\begin{align}
    \rBCS = \prod_{p=0}^M \left(u_p + v_p e^{i\lambda_p}\Cdag{p}\Cdag{ \bar{p}} \right) \rVac,
\end{align}
where $u_p$ and $v_p$ are real numbers such that $u_p^2 + v_p^2 = 1$, and $\lambda_p$ is a phase angle and is real valued. Define $\eta_p = \exp(i\lambda_p) v_p/u_p$, then we can write \cite{ring_nuclear_1980} 
\begin{subequations} \label{eq:BCS-our-formalism}
\begin{align}
    \rBCS &= \mathcal{N} \prod_{p=1}^M (1 + \eta_p \Pdag{p} ) \rVac \\
          &= \mathcal{N} \bigg(1 + \sum_p \eta_p \Pdag{p} + \sum_{p>q} \eta_p\eta_q \Pdag{p}\Pdag{q} + ... \nonumber\\ 
          &{} \quad \quad \quad + \eta_1\eta_2...\eta_M \Pdag{1}\Pdag{2}...\Pdag{M} \bigg) \rVac \label{eq:bcs_expanded}\\    
          &= \mathcal{N} \sum_{N=0}^{M} \frac{1}{N!} \left(\GamD\right)^N \rVac, \label{eq:bcs_as_sum_AGP}
\end{align}
\end{subequations}
where $\mathcal{N} = 1/\sqrt{\BCSnorm}$. One can readily see from \Eq{\ref{eq:bcs_as_sum_AGP}} and \Eq{\ref{eq:AGP_as_geminal}} that the BCS wavefunction is a superposition of AGPs with different numbers of pairs up to a normalization factor.

Implementation of $\rBCS$ on a quantum computer under the transformation \Eq{\ref{eqs:pair-qubit-mapping}} takes the form
\begin{align} \label{eq:BCS_qc_exapnded}
    \rBCS &= \mathcal{N} \bigg(1 + \sum_p \eta_p \Su{p} + \sum_{p>q} \eta_p\eta_q \Su{p}\Su{q} + ... \nonumber\\ 
    &{} \quad \quad \quad + \eta_1...\eta_M \Su{M}...\Su{2}\Su{1} \bigg) \rsVac.
\end{align}
We show that this can be implemented efficiently with a depth of \bigO{1} using single-qubit rotations. Explicitly, define $\theta_p = 2\arctan({v_p/u_p})$, then the BCS state can be obtained by
\begin{align}
    \rBCS = \prod_{p=1}^M e^{-i\lambda_p Z_p/2} e^{-i\theta_pY_p/2} \rsVac.
\end{align}
where $\exp(-i\lambda_p Z_p/2) \exp(-i\theta_p Y_p/2) = R_z(\lambda_p)R_y(\theta_p) = u(\lambda_p, \theta_p)$ is an elementary gate acting on qubit $p$. \cite{barenco_elementary_1995} This is because
\begin{subequations}\label{eqs:full-BCS}
\begin{align}
    \prod_{p=1}^M & e^{-i\lambda_p Z_p/2}  e^{-i\theta_pY_p/2} \rsVac = \nonumber \\
    &\prod_{p=1}^M e^{-i\lambda_p Z_p/2}  \left(\cos{\left(\frac{\theta_p}{2}\right)} I + (\Su{p} - \Sd{p}) \sin{\left(\frac{\theta_p}{2}\right)} \right)\rsVac \\
    &= \mathcal{N} \prod_{p=1}^M  \left(I + \tan{\left(\frac{\theta_p}{2}\right)} e^{i\lambda_p} \Su{p}\right)\rsVac \\
    &= \mathcal{N} \bigg(1 + \sum_p \eta_p \Su{p} + \sum_{p>q} \eta_p\eta_q \Su{p}\Su{q} + ... \nonumber\\ 
    &{} \quad \quad \quad + \eta_1...\eta_M \Su{M}...\Su{2}\Su{1} \bigg) \rsVac    
\end{align} 
\end{subequations}
which is the same as \Eq{\ref{eq:BCS_qc_exapnded}} up to an inconsequential global phase.
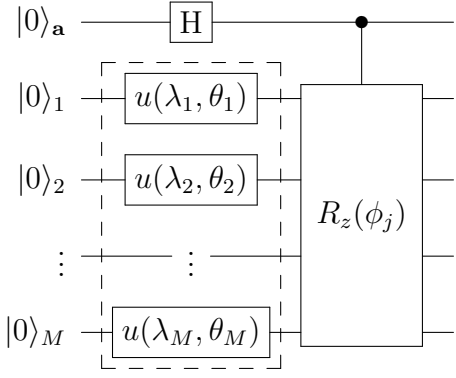
\begin{figure}[t]
    \leavevmode
    \centering
    \large
    \Qcircuit @C=1em @R=1em {
        \lstick{\ket{0}_\mathbf{a}} & \gate{\text{H}} & \ctrl{1} & \qw \\
        \lstick{\ket{0}_1} & \gate{u(\lambda_1,\theta_1)} & \multigate{3}{R_z(\phi_j)} & \qw \\
        \lstick{\ket{0}_2} & \gate{u(\lambda_2,\theta_2)} & \ghost{R_z(\phi_j)} & \qw \\
        \lstick{\vdots} & \push{\hspace{.2cm} \vdots \hspace{.2cm}} \qw & \ghost{R_z(\phi_j)} & \qw \\
        \lstick{\ket{0}_M} & \gate{u(\lambda_M,\theta_M)} & \ghost{R_z(\phi_j)} & \qw
        \gategroup{2}{2}{5}{2}{.7em}{--}
    }        
    \caption{Circuit for number projecting the BCS wavefunction. The operations in the dashed box implement the BCS state where u$(\lambda_p,\theta_p) = R_z(\lambda_p)R_y(\theta_p)$. The top wire corresponds to the ancilla qubit, and the controlled block consists of at most $M$ controlled $R_z(\phi_j)$.}
    \label{fig:prep_AGP}
\end{figure}

%---------------------------------------------------------------------
\subsection{Number projecting the BCS wavefunction} \label{sub:NP_AGP}
As it is clear from \Eq{\ref{eq:bcs_as_sum_AGP}}, AGP is nothing but the BCS state projected onto a particular number mode. Therefore, denote the number projection operator onto the $N$-pairs mode as \cite{ring_nuclear_1980, izmaylov_construction_2019}
\begin{align}
    \Proj = \frac{1}{2\pi} \int_0^{2\pi} d\phi \: e^{i\phi(\hat{N} - N)},
\end{align}
where $N$ is the desired number of pairs, and $\hat{N} = \frac{1}{2} \sum_p \N{p}$ in our formalism. Therefore, for the AGP with $N$ pairs we have 
\begin{align} \label{eq:PBCS}
    \rAGP = \frac{1}{\mathcal{N}}\Proj \rBCS.
\end{align}
Numerically, the integral in \Eq{\ref{eq:PBCS}} can be evaluated exactly by a sequence of ``half-projections" \cite{smeyers_half-projected_1973, yen_exact_2019, mihalka_half-projection_2020}
\begin{align} \label{eq:seq-half-projection}
    \Proj \rBCS = \prod_{j = 0}^{k} \frac{1}{2} \left(1 + e^{i\phi_j(\hat{N} - N)} \right)\rBCS,
\end{align}
where $\phi_j = \pi2^{-j}$ and $k = \floor{\log_2{ \text{max}\{N, M-N\}}}$. That is, every product eliminates half of the contaminants successively until the desired number of pairs is achieved. 

To implement the number projection on a quantum computer, first observe that 
\begin{align}
    e^{i\phi(\hat{N} - N)} \mapsto \gamma(\phi) \prod_{p=1}^{M}e^{-i \phi Z_p/2},
\end{align}
where $\exp\left({-i \frac{\phi}{2} Z_p}\right) = R_z(\phi)$ is an elementary gate acting on qubit $p$, and $\gamma(\phi) = \exp\left({i\phi(M/2 - N)}\right)$ is a global phase. On an ideal quantum computer, one can in principle prepare AGP by directly implementing \Eq{\ref{eq:seq-half-projection}} and using the methods prescribed in \Reference{\cite{childs_hamiltonian_2012}}. However, this would require a relatively deep circuit which could render its applicability on NISQ devices. Recognizing that for all practical purposes we want to compute expectation values of observables over AGP, a more efficient method is to trade depth for more measurements. This can be done by expanding the product in \Eq{\ref{eq:seq-half-projection}} into a sum and using \Eq{\ref{eq:PBCS}} to obtain
\begin{align} \label{eq:ev_A_agp_normalized}
    \frac{\lAGP \hat{A} \rAGP}{\AGPnorm} =
    \frac{ \sum_j \lBCS \hat{A} e^{i\phi_j(\hat{N} - N)}\rBCS}{ \sum_j \lBCS e^{i\phi_j(\hat{N} - N)} \rBCS},
\end{align}
where $\hat{A}$ is an arbitrary number-preserving observable and we used the property $\E{\Proj^{\dagger} \hat{A} \Proj}= \E{\hat{A} \Proj}$. If $\hat{A}$  is not number preserving or if it breaks seniority, then its expectation value over AGP is zero. Recently, \Reference{\cite{tsuchimochi_exact_2020}} used a similar technique to implement spin projection.

To minimize the number of measurements on a quantum computer, it is sufficient to only measure the terms in the numerator of the right hand side of \Eq{\ref{eq:ev_A_agp_normalized}}. The denominator is nothing but $\mathcal{N}^{2}\AGPnorm$ which can be computed on a classical computer efficiently; $\AGPnorm$ is an elementary symmetric polynomial over $\{\eta_p^2\}$, \cite{khamoshi_efficient_2019} and $\mathcal{N} = \prod_p u_p$ from \Eq{\ref{eq:BCS-our-formalism}}. Thus, using the expanded form of \Eq{\ref{eqs:full-BCS}} in particular, we obtain
\begin{align} \label{eq:EV_A_qc}
    \frac{\lAGP \hat{A} \rAGP}{\AGPnorm} = \mathcal{C} \sum_{j=1}^{n} \lBCS \hat{A} e^{i\phi_j(\hat{N} - N)}\rBCS,
\end{align}
where $n= 2^{k+1}$, $\phi_j = 2\pi(j-1)/n$ and
\begin{align}
    \mathcal{C} = \frac{1}{n} \frac{\BCSnorm}{\AGPnorm},
\end{align} 
is computed on a classical device. Furthermore, for numerical convenience we can scale the geminal coefficients beforehand, say by
\begin{align}\label{eq:normalize_eta}
    \eta_p \rightarrow \frac{\eta_p}{\sqrt[2N]{\AGPnorm}},
\end{align}
so that $\AGPnorm = 1$. 

Lastly, since $\hat{A}e^{i\phi_j(\hat{N} - N)}$ is not Hermitian, we comment on how to evaluate each term in the right hand side of \Eq{\ref{eq:EV_A_qc}} on a quantum computer. For this, we invoke the well-known Hadamard test \cite{aharonov_polynomial_2009,stair_multireference_2020} for which we must introduce a single ancilla qubit and use 2-qubit controlled-$R_z$ rotations to obtain
\begin{align}\label{eq:Hadamard}
    \ket{\Psi} = \frac{1}{\sqrt{2}} \left(\rBCS\ket{0}_{\mathbf{a}} + \ProjOp{\phi_j}\rBCS\ket{1}_{\mathbf{a}}\right),
\end{align}
where $\ProjOp{\phi_j} = \bigotimes_{p=1}^M R_z(\phi_j)$. See \Fig{\ref{fig:prep_AGP}} for a circuit diagram. Then for an observable $\hat{A}$, we measure $\text{Re}\left[2\E{\Sd{\mathbf{a}}}\right] = \E{X_{\mathbf{a}}}$ for each $\phi_j$. Note that it is not necessary to involve $\hat{A}$ in the Hadamard test; since $\hat{A}$ is Hermitian, we can always measure the qubits in the basis where $\hat{A}$ is diagonal, hence avoiding additional controlled operations. For example, for a Hamiltonian written as a sum of Pauli ``words", $\hat{A}$ is simply a product of Pauli matrices for which the change of basis is straightforward.

%---------------------------------------------------------------------
\subsection{Correlating AGP with the unitary pair-hopper on a quantum computer}\label{sub:pair-hoppers}
Expectation values of many-body operators over AGP can be efficiently computed on a classical computer. \cite{khamoshi_efficient_2019} The advantage of using a quantum computer emerges when we want to correlate AGP by some unitary operator. In our method, adding correlation by a unitary operator $\hat{U}$ simply amounts to changing $\hat{A} \rightarrow \hat{U}^{\dagger}\hat{A}\hat{U}$ in \Eq{\ref{eq:EV_A_qc}} while everything else remains the same.

There are several candidates for $\hat{U}$. For example, \Reference{\cite{matsuzawa_jastrow-type_2020}} suggested using a Jastrow-type correlator applied to AGP. In this paper, we propose to use another correlator which we derive from the 2-body killer of AGP, i.e. $\Killer{pq}\rAGP = 0$. \cite{henderson_geminal-based_2019} It was shown that one can build excitations on top of AGP using the adjoint of the killer, which can be written as 
\begin{align}
    \KillerAd{pq} = \eta_q^2 \Pdag{p}\Pp{q} + \eta_p^2 \Pdag{q}\Pp{p} + \frac{1}{2} \eta_p\eta_q \left( \N{p}\N{q} - \N{p} - \N{q} \right).
\end{align}
Here, we propose to build an anti-Hermitian correlator from $\Killer{pq} - \KillerAd{pq} \propto \Pdag{p}\Pp{q} - \Pdag{q}\Pp{p}$, and use its exponential as a unitary correlator. Formally, define the anti-Hermitian pair-hopper operator as
\begin{align}
    \PHoppers = \sum_{p>q}^M \tau_{pq} \left( \Pdag{p}\Pp{q} - \Pdag{q}\Pp{p} \right),
\end{align}
where $\tau_{pq}$ are the variational coefficients and $\tau_{pq} = -\tau_{qp}$. Under the transformation in \Eq{\ref{eqs:pair-qubit-mapping}}, each term becomes
\begin{align}
    \left(\Pdag{p}\Pp{q} - \Pdag{q}\Pp{p}\right) \mapsto i \left(X_pY_q - Y_pX_q \right)/2.
\end{align}

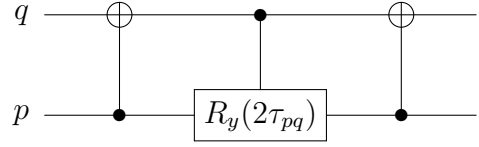
\begin{figure}[t] 
    \leavevmode
    \centering
    \large
    \Qcircuit @C=2em @R=2em {
        \lstick{q} & \targ & \ctrl{1} & \targ & \qw \\
        \lstick{p} & \ctrl{-1} &  \gate{R_y(2\tau_{pq})} & \ctrl{-1} & \qw 
    }        
    \caption{Circuit for implementing $\hat{U}_{pq}$. The top and bottom wires correspond to the qubits indexed by $q$ and $p$ respectively. $R_y(2\tau_{pq}) = \exp{(-\tau_{pq}Y)}$ is an elementary gate.}
    \label{fig:pair-hoppers}
\end{figure}

We define the unitary pair-hopper correlator $\hat{U}$ as
\begin{align} \label{eq:U-PairHoppers}
    \mathlarger{\hat{U}} \equiv \exp(\PHoppers) = \exp{\left( i \sum_{p>q} \frac{\tau_{pq}}{2} \left(X_pY_q - Y_pX_q \right) \right)}.
\end{align}
In order to implement \Eq{\ref{eq:U-PairHoppers}} on a quantum computer, we decompose $\hat{U}$ as a product of elementary gates. However, since the terms in the anti-Hermitian pair-hopper do not commute, we resort to implementing $\hat{U}$ as a product of exponentials, that is
\begin{align} \label{eq:U-Trott-PairHoppers}
    \mathlarger{\hat{U}} = \prod_{p>q}^M \exp{ \left( i \frac{\tau_{pq}}{2} \left(X_pY_q - Y_pX_q \right) \right)}.
\end{align}
Notice that the matrix representation of each exponential is simply
\begin{equation} \label{eq:Upq_matrix}
    \mathlarger{\hat{U}}_{pq} = 
    \begin{pmatrix}
        1 & 0 & 0 & 0 \\ 
        0 & \cos{\tau_{pq}} &-\sin{\tau_{pq}} & 0 \\
        0 & \sin{\tau_{pq}} & \cos{\tau_{pq}} & 0 \\
        0 & 0 & 0 & 1 \\ 
    \end{pmatrix},
\end{equation}
where $\hat{U}_{pq} \equiv \exp{\left(i\tau_{pq}(X_pY_q - Y_pX_q)/2\right)}$. As such, $\hat{U}_{pq}$ can be conceived of as a 2-qubit entangling gate \cite{egger_entanglement_2019, barkoutsos_quantum_2018, gard_efficient_2020, barron_preserving_2020} leaving the states $\ket{00}$ and $\ket{11}$ untouched, and 
\begin{subequations}
\begin{align}
    \ket{01} &\rightarrow \cos{(\tau_{pq})}\ket{01} + \sin{(\tau_{pq})} \ket{10}, \\ 
    \ket{10} &\rightarrow \cos{(\tau_{pq})}\ket{10} - \sin{(\tau_{pq})} \ket{01}.
\end{align}
\end{subequations}
Note that $\hat{U}_{pq}$ is number conserving. \Fig{\ref{fig:pair-hoppers}} shows a simple circuit to implement $\hat{U}_{pq}$. \cite{barkoutsos_quantum_2018}

%====================================================================%
%                      ** APPLICATION **                             %
%====================================================================%
\section{Application} \label{sec:application}
In this section, we benchmark the unitary pair-hopper ansatz on the reduced BCS Hamiltonian. Using \Eq{\ref{eqs:pair-qubit-mapping}}, the Hamiltonian in \Eq{\ref{eq:pairing_Ham_fermions}} can be written as
\begin{align}
    \Ham = \sum_{p = 1}^M (\epsilon_p - \frac{G}{2})(1 - Z_p)  - \frac{G}{2} \sum_{p>q}^M \left( X_pX_q + Y_pY_q\right),
\end{align}
The geminal coefficients can be chosen to be real, and we normalize $\eta_p$ according to \Eq{\ref{eq:normalize_eta}} so that $\AGPnorm = 1$. Our goal is to obtain the ground state by variationally optimizing the energy as follows
\begin{align}
    E_{\text{gs}} = \underset{ \tau }{\text{min}} \left\{ \frac{\Eagp{\hat{U}^{\dagger}(\tau) \hat{H} \hat{U}(\tau)}}{\AGPnorm} \right\}.
\end{align}
The gradients are computed numerically when necessary. Since the anti-Hermitian pair-hopper does not form a closed and compact subalgebra, the ansatz is generally order-dependent. \cite{evangelista_exact_2019, grimsley_is_2020, izmaylov_order_2020} Therefore, for simplicity, we take the following ordering 
\begin{align}
    \prod_{p>q}^M \hat{U}_{pq} = \hat{U}_{MM-1}...\hat{U}_{32}\hat{U}_{M1}...\hat{U}_{31}\hat{U}_{21}.
\end{align}
\begin{figure*}[t]%
    \centering
    \subfloat{{\includegraphics[width=9cm]{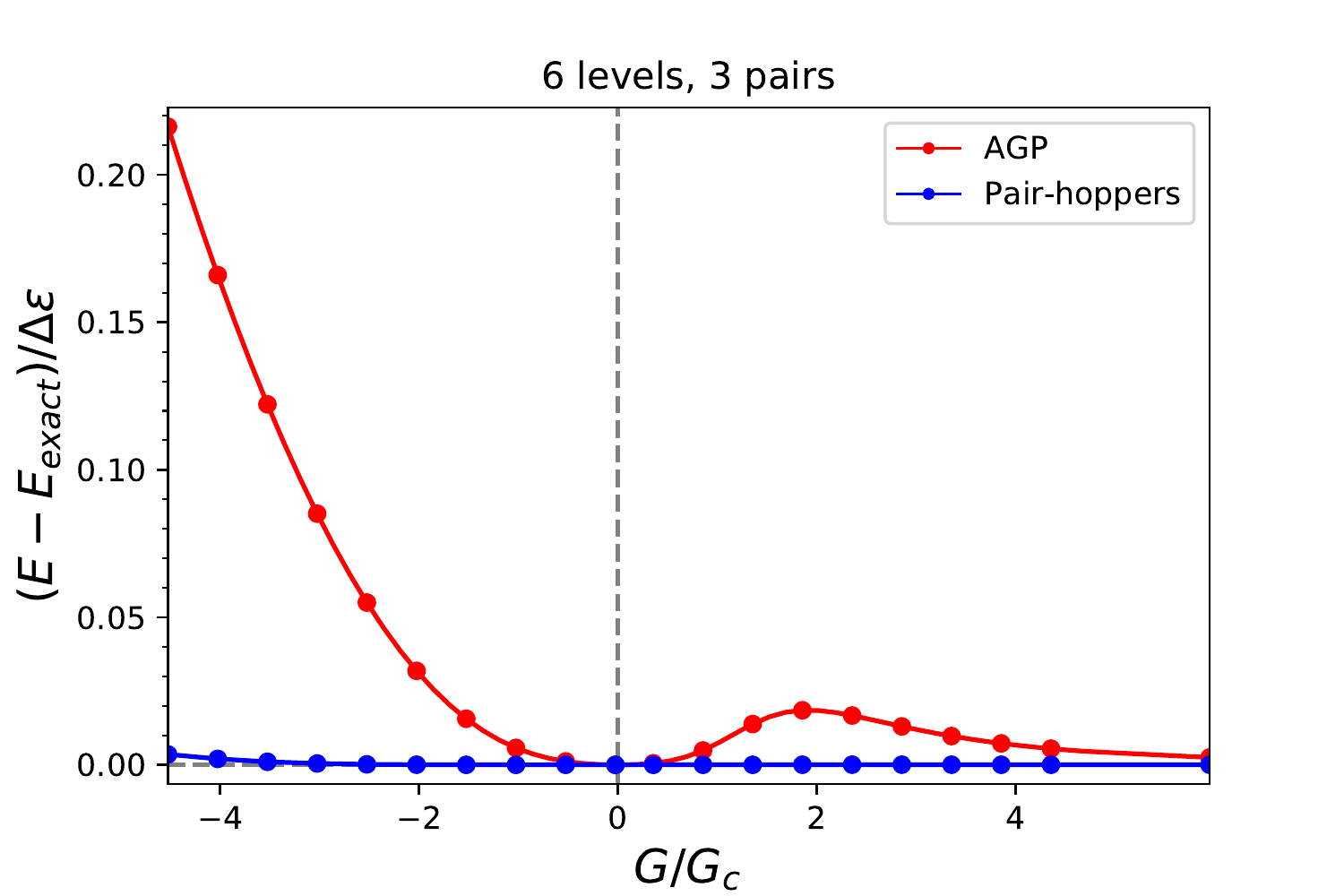}}}%
    \subfloat{{\includegraphics[width=9cm]{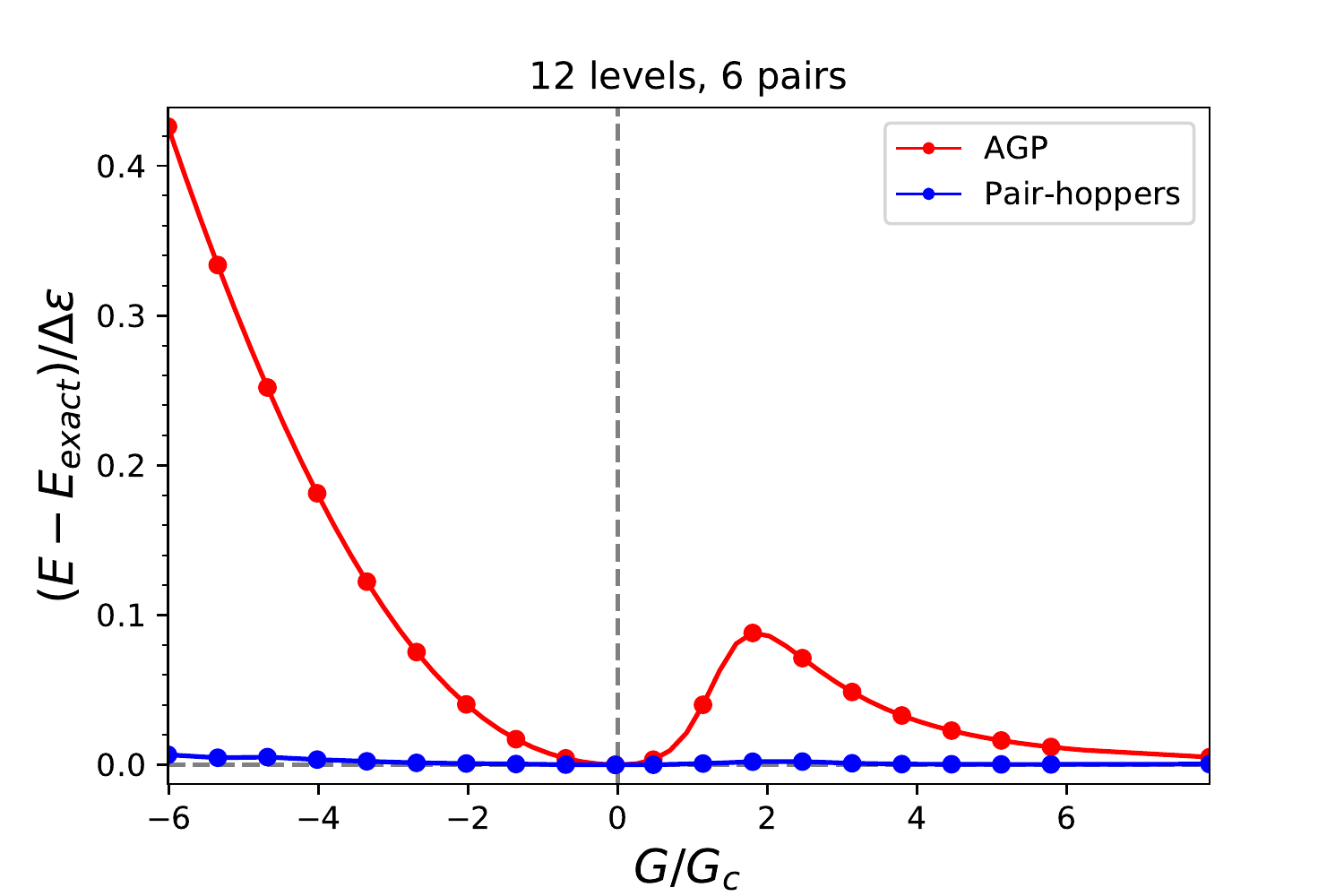}}}%
    \caption{Energy error of the unitary pair-hopper ansatz  optimized for the pairing Hamiltonian as a function of $G/G_c$. The left figure corresponds to $6$ levels and the right is for $12$ levels at half-filing}%
    \label{fig:errE}
\end{figure*}
We observe that in the limit where the dimension of this ansatz is equal to that of FCI (DOCI), i.e. when $M = 4$ and  $N = 2$, the energy converges to the exact value. While this justifies the use of this particular ansatz, \cite{evangelista_exact_2019} finding an optimal ordering for pair-hoppers will be the subject of future research. For this choice, the depth and number of CNOTs grow asymptotically as \bigO{M^2} on an all-to-all connected quantum computer; on a realistic quantum computer with limited links, the scaling might be different but is no worse than \bigO{M^3} \cite{tannu_not_2019}. On any architecture, the circuit is to be transpiled in such a way that maximizes parallel executions without changing the ordering. The number of variational parameters is $M(M-1)/2$.

The numerical simulations reported in this section were performed on a classical computer in the absence of noise. We used IBM's Qiskit libraries \cite{noauthor_qiskit_2019} for the backend simulator. The unconstrained minimization of energy were carried out using a combination of Sequential Least Squares Programming (SLSQP) and limited-memory Broyden–Fletcher–Goldfarb–Shanno (L-BFGS) algorithms \cite{kraft_software_1988, byrd_limited_1995}. We report our numerical simulations for $M=6, 12$. The circuit depth associated with each system is 19 and 67 respectively, and the number of two-qubit gates (CNOTs) is $M+3M(M-1)/2$ for any system of size $M$. The transpilation of all circuits is set to default throughout our calculations and the underlying architecture is assumed to be all-to-all.

\begin{figure*}[t]%
    \centering
    \subfloat{{\includegraphics[width=9cm]{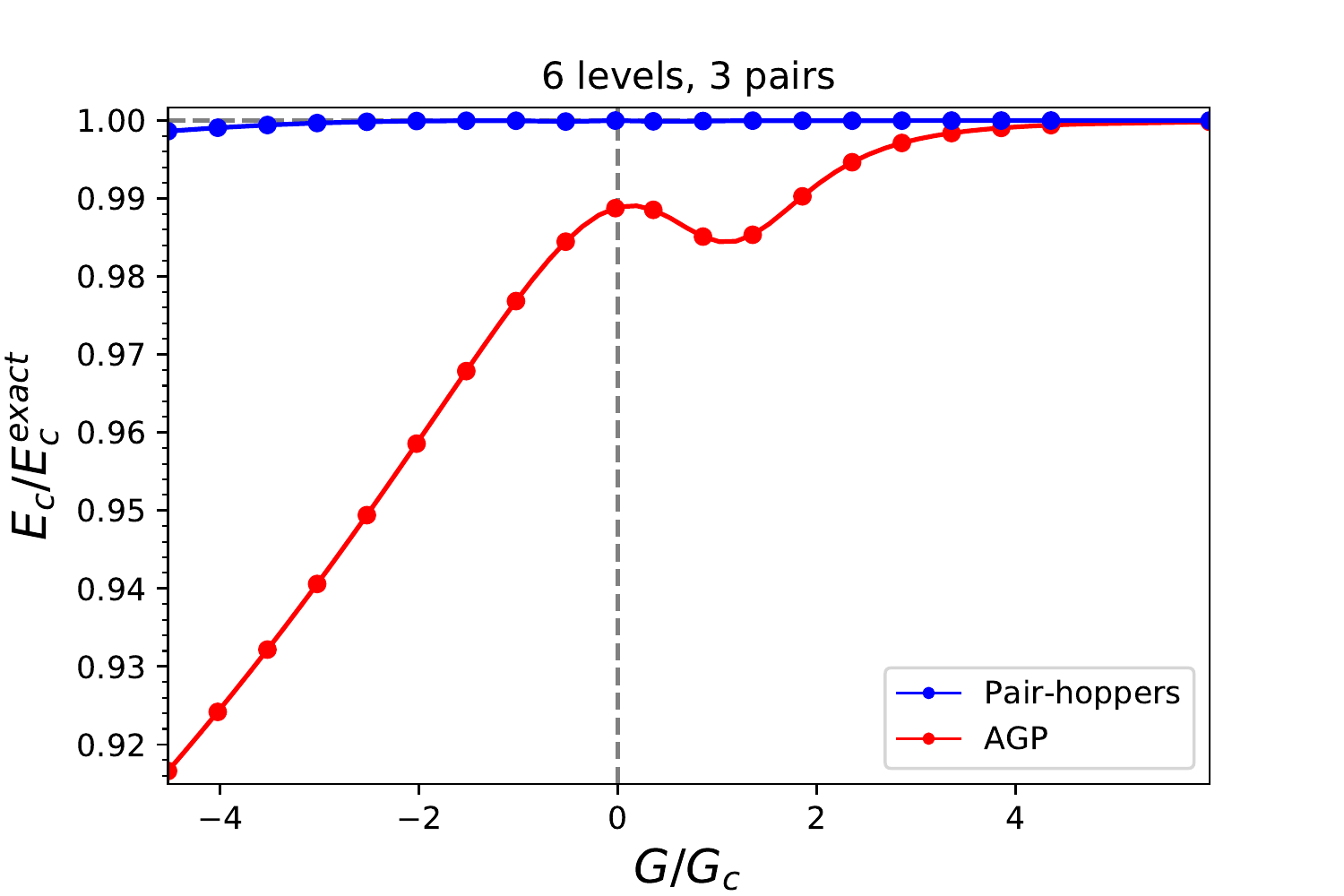}}}%
    \subfloat{{\includegraphics[width=9cm]{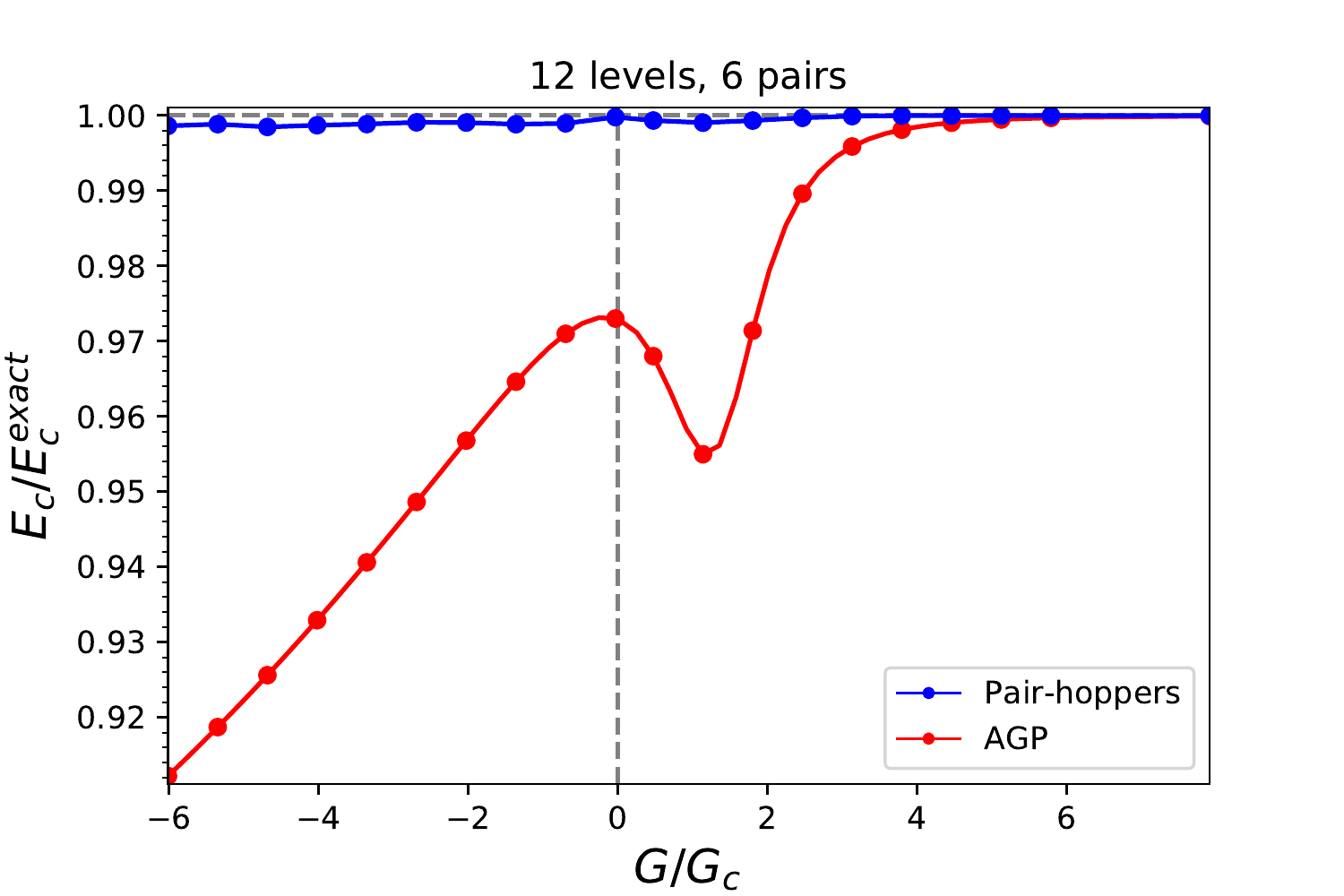}}}%
    \caption{Recovered correlation energy as a fraction of the exact correlation. The left figure corresponds to $6$ levels and the right is for $12$ levels at half-filing}%
    \label{fig:Ecorr}%
\end{figure*}

\Fig{\ref{fig:errE}} shows the energy error for the pairing Hamiltonian as a function of $G/G_c$ at half-filling. As discussed in the introduction, although the attractive side has been the most  challenging regime for most conventional methods, we demonstrate our method in both the attractive and repulsive regimes for completeness. We have chosen half-filling because it corresponds to the case where AGP contains the largest number of Slater determinants. Nevertheless, the complexity of our circuit and the asymptotic cost of our method is agnostic to the number of pairs. As demonstrated in the figure, the bare AGP wavefunction is exact on the attractive side in the limit where $G/G_c \rightarrow +\infty$, but on the repulsive side, $G/G_c<0$, it gradually deteriorates as $G$ gets smaller. In contrast, the application of unitary pair-hoppers on AGP recovers a significant portion of the correlation energy in both the attractive and repulsive regimes. Although the optimized energies are not exact, this shows that our ansatz is robust and highly accurate for the pairing Hamiltonian. To make this point clear, we also plot the fraction of the correlation energy computed as 
\begin{align}
    E_{c}/E_{c}^{\text{exact}} = \frac{E - E_{\text{HF}}}{E_{\text{exact}} - E_{\text{HF}}},
\end{align}
as a function of $G/G_c$ in \Fig{\ref{fig:Ecorr}}. The plots show that more than 99\% of the correlation energy has been recovered by the unitary pair-hopper in either system. 

As was noted earlier in this paper and elsewhere \cite{henderson_geminal-based_2019,henderson_correlating_2020}, the interaction in the pairing Hamiltonian is infinite-range which makes the exact energy a nonlinear function of system size. This complicates the discussion about size-extensivity of the ansatz which we do not intend to address in this work.

%====================================================================%
%                      ** CONCLUSIONS **                             %
%====================================================================%
\section{Conclusions}
Motivated by using AGP as an initial reference for strongly correlated systems, we have shown an efficient and NISQ-friendly implementation of AGP on a quantum computer. The main purpose of so doing is to build unitary ansatze over AGP in order capture the remaining correlation error. Such ansatze are not easily accessible on a classical computer, but they can be implemented efficiently on a quantum computer as we demonstrated in this paper.  

Our method for implementing AGP uses a hybrid of classical and quantum methods wherein we first optimize the geminal coefficients on a classical computer then use them to implement AGP on a quantum device. The quantum implementation involves two steps: First, we implement the BCS wavefunction efficiently using single-qubit rotations with the total depth of \bigO{1}. Second, we carry out the number projection with at most \bigO{M} measurements. The depth and number of CNOTs involved in the number projection circuit grow linearly in system size. A key observation that contributes to a resource-friendly implementation of AGP and its correlators is the use of an economic mapping between fermion pairs and qubits. Under this mapping, the Pauli Z strings associated with the fermionic anticommutation are inherently absent, and, compared to an implementation based on fermionic representation, our approach uses half the number of qubits thus leading to shallower circuits.

Finally, we demonstrated how to correlate AGP on a quantum computer by applying the unitary pair-hopper on AGP. Our numerical simulations show that this ansatz is capable of recovering highly accurate energies in both the attractive and repulsive regimes of the pairing Hamiltonian, thereby making it a potentially useful ansatz for more sophisticated Hamiltonians.

%====================================================================%
%                    ** ACKNOWLEDGMENTS **                           %
%====================================================================%
\section{Acknowledgments} 
The authors were supported by the U.S. Department of Energy
under Award No. DE-SC0019374. G.E.S. is a Welch Foundation
Chair (C-0036). We thank Thomas Henderson for valuable discussions in regards to number-projection. A.K. is thankful to Nicholas Stair for helpful discussions about the circuit implementation.

%====================================================================%
%                      ** BIBLIOGRAPHY **                            %
%====================================================================%
\bibliography{main}

\begin{thebibliography}{81}%
\makeatletter
\providecommand \@ifxundefined [1]{%
 \@ifx{#1\undefined}
}%
\providecommand \@ifnum [1]{%
 \ifnum #1\expandafter \@firstoftwo
 \else \expandafter \@secondoftwo
 \fi
}%
\providecommand \@ifx [1]{%
 \ifx #1\expandafter \@firstoftwo
 \else \expandafter \@secondoftwo
 \fi
}%
\providecommand \natexlab [1]{#1}%
\providecommand \enquote  [1]{``#1''}%
\providecommand \bibnamefont  [1]{#1}%
\providecommand \bibfnamefont [1]{#1}%
\providecommand \citenamefont [1]{#1}%
\providecommand \href@noop [0]{\@secondoftwo}%
\providecommand \href [0]{\begingroup \@sanitize@url \@href}%
\providecommand \@href[1]{\@@startlink{#1}\@@href}%
\providecommand \@@href[1]{\endgroup#1\@@endlink}%
\providecommand \@sanitize@url [0]{\catcode `\\12\catcode `\$12\catcode
  `\&12\catcode `\#12\catcode `\^12\catcode `\_12\catcode `\%12\relax}%
\providecommand \@@startlink[1]{}%
\providecommand \@@endlink[0]{}%
\providecommand \url  [0]{\begingroup\@sanitize@url \@url }%
\providecommand \@url [1]{\endgroup\@href {#1}{\urlprefix }}%
\providecommand \urlprefix  [0]{URL }%
\providecommand \Eprint [0]{\href }%
\providecommand \doibase [0]{http://dx.doi.org/}%
\providecommand \selectlanguage [0]{\@gobble}%
\providecommand \bibinfo  [0]{\@secondoftwo}%
\providecommand \bibfield  [0]{\@secondoftwo}%
\providecommand \translation [1]{[#1]}%
\providecommand \BibitemOpen [0]{}%
\providecommand \bibitemStop [0]{}%
\providecommand \bibitemNoStop [0]{.\EOS\space}%
\providecommand \EOS [0]{\spacefactor3000\relax}%
\providecommand \BibitemShut  [1]{\csname bibitem#1\endcsname}%
\let\auto@bib@innerbib\@empty
%</preamble>
\bibitem [{\citenamefont {Babbush}\ \emph {et~al.}(2018)\citenamefont
  {Babbush}, \citenamefont {Wiebe}, \citenamefont {McClean}, \citenamefont
  {McClain}, \citenamefont {Neven},\ and\ \citenamefont
  {Chan}}]{babbush_low-depth_2018}%
  \BibitemOpen
  \bibfield  {author} {\bibinfo {author} {\bibfnamefont {R.}~\bibnamefont
  {Babbush}}, \bibinfo {author} {\bibfnamefont {N.}~\bibnamefont {Wiebe}},
  \bibinfo {author} {\bibfnamefont {J.}~\bibnamefont {McClean}}, \bibinfo
  {author} {\bibfnamefont {J.}~\bibnamefont {McClain}}, \bibinfo {author}
  {\bibfnamefont {H.}~\bibnamefont {Neven}}, \ and\ \bibinfo {author}
  {\bibfnamefont {G.~K.-L.}\ \bibnamefont {Chan}},\ }\href {\doibase
  10.1103/PhysRevX.8.011044} {\bibfield  {journal} {\bibinfo  {journal} {Phys.
  Rev. X}\ }\textbf {\bibinfo {volume} {8}},\ \bibinfo {pages} {011044}
  (\bibinfo {year} {2018})}\BibitemShut {NoStop}%
\bibitem [{\citenamefont {Cao}\ \emph {et~al.}(2019)\citenamefont {Cao},
  \citenamefont {Romero}, \citenamefont {Olson}, \citenamefont {Degroote},
  \citenamefont {Johnson}, \citenamefont {Kieferová}, \citenamefont
  {Kivlichan}, \citenamefont {Menke}, \citenamefont {Peropadre}, \citenamefont
  {Sawaya}, \citenamefont {Sim}, \citenamefont {Veis},\ and\ \citenamefont
  {Aspuru-Guzik}}]{cao_quantum_2019}%
  \BibitemOpen
  \bibfield  {author} {\bibinfo {author} {\bibfnamefont {Y.}~\bibnamefont
  {Cao}}, \bibinfo {author} {\bibfnamefont {J.}~\bibnamefont {Romero}},
  \bibinfo {author} {\bibfnamefont {J.~P.}\ \bibnamefont {Olson}}, \bibinfo
  {author} {\bibfnamefont {M.}~\bibnamefont {Degroote}}, \bibinfo {author}
  {\bibfnamefont {P.~D.}\ \bibnamefont {Johnson}}, \bibinfo {author}
  {\bibfnamefont {M.}~\bibnamefont {Kieferová}}, \bibinfo {author}
  {\bibfnamefont {I.~D.}\ \bibnamefont {Kivlichan}}, \bibinfo {author}
  {\bibfnamefont {T.}~\bibnamefont {Menke}}, \bibinfo {author} {\bibfnamefont
  {B.}~\bibnamefont {Peropadre}}, \bibinfo {author} {\bibfnamefont {N.~P.~D.}\
  \bibnamefont {Sawaya}}, \bibinfo {author} {\bibfnamefont {S.}~\bibnamefont
  {Sim}}, \bibinfo {author} {\bibfnamefont {L.}~\bibnamefont {Veis}}, \ and\
  \bibinfo {author} {\bibfnamefont {A.}~\bibnamefont {Aspuru-Guzik}},\ }\href
  {\doibase 10.1021/acs.chemrev.8b00803} {\bibfield  {journal} {\bibinfo
  {journal} {Chem. Rev.}\ }\textbf {\bibinfo {volume} {119}},\ \bibinfo {pages}
  {10856} (\bibinfo {year} {2019})}\BibitemShut {NoStop}%
\bibitem [{\citenamefont {McArdle}\ \emph {et~al.}(2020)\citenamefont
  {McArdle}, \citenamefont {Endo}, \citenamefont {Aspuru-Guzik}, \citenamefont
  {Benjamin},\ and\ \citenamefont {Yuan}}]{mcardle_quantum_2020}%
  \BibitemOpen
  \bibfield  {author} {\bibinfo {author} {\bibfnamefont {S.}~\bibnamefont
  {McArdle}}, \bibinfo {author} {\bibfnamefont {S.}~\bibnamefont {Endo}},
  \bibinfo {author} {\bibfnamefont {A.}~\bibnamefont {Aspuru-Guzik}}, \bibinfo
  {author} {\bibfnamefont {S.~C.}\ \bibnamefont {Benjamin}}, \ and\ \bibinfo
  {author} {\bibfnamefont {X.}~\bibnamefont {Yuan}},\ }\href {\doibase
  10.1103/RevModPhys.92.015003} {\bibfield  {journal} {\bibinfo  {journal}
  {Rev. Mod. Phys.}\ }\textbf {\bibinfo {volume} {92}},\ \bibinfo {pages}
  {015003} (\bibinfo {year} {2020})}\BibitemShut {NoStop}%
\bibitem [{\citenamefont {Preskill}(2018)}]{preskill_quantum_2018}%
  \BibitemOpen
  \bibfield  {author} {\bibinfo {author} {\bibfnamefont {J.}~\bibnamefont
  {Preskill}},\ }\href {\doibase 10.22331/q-2018-08-06-79} {\bibfield
  {journal} {\bibinfo  {journal} {Quantum}\ }\textbf {\bibinfo {volume} {2}},\
  \bibinfo {pages} {79} (\bibinfo {year} {2018})},\ \bibinfo {note} {arXiv:
  1801.00862}\BibitemShut {NoStop}%
\bibitem [{\citenamefont {Arute}\ \emph {et~al.}(2019)\citenamefont {Arute},
  \citenamefont {Arya}, \citenamefont {Babbush}, \citenamefont {Bacon},
  \citenamefont {Bardin}, \citenamefont {Barends}, \citenamefont {Biswas},
  \citenamefont {Boixo}, \citenamefont {Brandao}, \citenamefont {Buell},
  \citenamefont {Burkett}, \citenamefont {Chen}, \citenamefont {Chen},
  \citenamefont {Chiaro}, \citenamefont {Collins}, \citenamefont {Courtney},
  \citenamefont {Dunsworth}, \citenamefont {Farhi}, \citenamefont {Foxen},
  \citenamefont {Fowler}, \citenamefont {Gidney}, \citenamefont {Giustina},
  \citenamefont {Graff}, \citenamefont {Guerin}, \citenamefont {Habegger},
  \citenamefont {Harrigan}, \citenamefont {Hartmann}, \citenamefont {Ho},
  \citenamefont {Hoffmann}, \citenamefont {Huang}, \citenamefont {Humble},
  \citenamefont {Isakov}, \citenamefont {Jeffrey}, \citenamefont {Jiang},
  \citenamefont {Kafri}, \citenamefont {Kechedzhi}, \citenamefont {Kelly},
  \citenamefont {Klimov}, \citenamefont {Knysh}, \citenamefont {Korotkov},
  \citenamefont {Kostritsa}, \citenamefont {Landhuis}, \citenamefont
  {Lindmark}, \citenamefont {Lucero}, \citenamefont {Lyakh}, \citenamefont
  {Mandrà}, \citenamefont {McClean}, \citenamefont {McEwen}, \citenamefont
  {Megrant}, \citenamefont {Mi}, \citenamefont {Michielsen}, \citenamefont
  {Mohseni}, \citenamefont {Mutus}, \citenamefont {Naaman}, \citenamefont
  {Neeley}, \citenamefont {Neill}, \citenamefont {Niu}, \citenamefont {Ostby},
  \citenamefont {Petukhov}, \citenamefont {Platt}, \citenamefont {Quintana},
  \citenamefont {Rieffel}, \citenamefont {Roushan}, \citenamefont {Rubin},
  \citenamefont {Sank}, \citenamefont {Satzinger}, \citenamefont {Smelyanskiy},
  \citenamefont {Sung}, \citenamefont {Trevithick}, \citenamefont
  {Vainsencher}, \citenamefont {Villalonga}, \citenamefont {White},
  \citenamefont {Yao}, \citenamefont {Yeh}, \citenamefont {Zalcman},
  \citenamefont {Neven},\ and\ \citenamefont {Martinis}}]{arute_quantum_2019}%
  \BibitemOpen
  \bibfield  {author} {\bibinfo {author} {\bibfnamefont {F.}~\bibnamefont
  {Arute}}, \bibinfo {author} {\bibfnamefont {K.}~\bibnamefont {Arya}},
  \bibinfo {author} {\bibfnamefont {R.}~\bibnamefont {Babbush}}, \bibinfo
  {author} {\bibfnamefont {D.}~\bibnamefont {Bacon}}, \bibinfo {author}
  {\bibfnamefont {J.~C.}\ \bibnamefont {Bardin}}, \bibinfo {author}
  {\bibfnamefont {R.}~\bibnamefont {Barends}}, \bibinfo {author} {\bibfnamefont
  {R.}~\bibnamefont {Biswas}}, \bibinfo {author} {\bibfnamefont
  {S.}~\bibnamefont {Boixo}}, \bibinfo {author} {\bibfnamefont {F.~G. S.~L.}\
  \bibnamefont {Brandao}}, \bibinfo {author} {\bibfnamefont {D.~A.}\
  \bibnamefont {Buell}}, \bibinfo {author} {\bibfnamefont {B.}~\bibnamefont
  {Burkett}}, \bibinfo {author} {\bibfnamefont {Y.}~\bibnamefont {Chen}},
  \bibinfo {author} {\bibfnamefont {Z.}~\bibnamefont {Chen}}, \bibinfo {author}
  {\bibfnamefont {B.}~\bibnamefont {Chiaro}}, \bibinfo {author} {\bibfnamefont
  {R.}~\bibnamefont {Collins}}, \bibinfo {author} {\bibfnamefont
  {W.}~\bibnamefont {Courtney}}, \bibinfo {author} {\bibfnamefont
  {A.}~\bibnamefont {Dunsworth}}, \bibinfo {author} {\bibfnamefont
  {E.}~\bibnamefont {Farhi}}, \bibinfo {author} {\bibfnamefont
  {B.}~\bibnamefont {Foxen}}, \bibinfo {author} {\bibfnamefont
  {A.}~\bibnamefont {Fowler}}, \bibinfo {author} {\bibfnamefont
  {C.}~\bibnamefont {Gidney}}, \bibinfo {author} {\bibfnamefont
  {M.}~\bibnamefont {Giustina}}, \bibinfo {author} {\bibfnamefont
  {R.}~\bibnamefont {Graff}}, \bibinfo {author} {\bibfnamefont
  {K.}~\bibnamefont {Guerin}}, \bibinfo {author} {\bibfnamefont
  {S.}~\bibnamefont {Habegger}}, \bibinfo {author} {\bibfnamefont {M.~P.}\
  \bibnamefont {Harrigan}}, \bibinfo {author} {\bibfnamefont {M.~J.}\
  \bibnamefont {Hartmann}}, \bibinfo {author} {\bibfnamefont {A.}~\bibnamefont
  {Ho}}, \bibinfo {author} {\bibfnamefont {M.}~\bibnamefont {Hoffmann}},
  \bibinfo {author} {\bibfnamefont {T.}~\bibnamefont {Huang}}, \bibinfo
  {author} {\bibfnamefont {T.~S.}\ \bibnamefont {Humble}}, \bibinfo {author}
  {\bibfnamefont {S.~V.}\ \bibnamefont {Isakov}}, \bibinfo {author}
  {\bibfnamefont {E.}~\bibnamefont {Jeffrey}}, \bibinfo {author} {\bibfnamefont
  {Z.}~\bibnamefont {Jiang}}, \bibinfo {author} {\bibfnamefont
  {D.}~\bibnamefont {Kafri}}, \bibinfo {author} {\bibfnamefont
  {K.}~\bibnamefont {Kechedzhi}}, \bibinfo {author} {\bibfnamefont
  {J.}~\bibnamefont {Kelly}}, \bibinfo {author} {\bibfnamefont {P.~V.}\
  \bibnamefont {Klimov}}, \bibinfo {author} {\bibfnamefont {S.}~\bibnamefont
  {Knysh}}, \bibinfo {author} {\bibfnamefont {A.}~\bibnamefont {Korotkov}},
  \bibinfo {author} {\bibfnamefont {F.}~\bibnamefont {Kostritsa}}, \bibinfo
  {author} {\bibfnamefont {D.}~\bibnamefont {Landhuis}}, \bibinfo {author}
  {\bibfnamefont {M.}~\bibnamefont {Lindmark}}, \bibinfo {author}
  {\bibfnamefont {E.}~\bibnamefont {Lucero}}, \bibinfo {author} {\bibfnamefont
  {D.}~\bibnamefont {Lyakh}}, \bibinfo {author} {\bibfnamefont
  {S.}~\bibnamefont {Mandrà}}, \bibinfo {author} {\bibfnamefont {J.~R.}\
  \bibnamefont {McClean}}, \bibinfo {author} {\bibfnamefont {M.}~\bibnamefont
  {McEwen}}, \bibinfo {author} {\bibfnamefont {A.}~\bibnamefont {Megrant}},
  \bibinfo {author} {\bibfnamefont {X.}~\bibnamefont {Mi}}, \bibinfo {author}
  {\bibfnamefont {K.}~\bibnamefont {Michielsen}}, \bibinfo {author}
  {\bibfnamefont {M.}~\bibnamefont {Mohseni}}, \bibinfo {author} {\bibfnamefont
  {J.}~\bibnamefont {Mutus}}, \bibinfo {author} {\bibfnamefont
  {O.}~\bibnamefont {Naaman}}, \bibinfo {author} {\bibfnamefont
  {M.}~\bibnamefont {Neeley}}, \bibinfo {author} {\bibfnamefont
  {C.}~\bibnamefont {Neill}}, \bibinfo {author} {\bibfnamefont {M.~Y.}\
  \bibnamefont {Niu}}, \bibinfo {author} {\bibfnamefont {E.}~\bibnamefont
  {Ostby}}, \bibinfo {author} {\bibfnamefont {A.}~\bibnamefont {Petukhov}},
  \bibinfo {author} {\bibfnamefont {J.~C.}\ \bibnamefont {Platt}}, \bibinfo
  {author} {\bibfnamefont {C.}~\bibnamefont {Quintana}}, \bibinfo {author}
  {\bibfnamefont {E.~G.}\ \bibnamefont {Rieffel}}, \bibinfo {author}
  {\bibfnamefont {P.}~\bibnamefont {Roushan}}, \bibinfo {author} {\bibfnamefont
  {N.~C.}\ \bibnamefont {Rubin}}, \bibinfo {author} {\bibfnamefont
  {D.}~\bibnamefont {Sank}}, \bibinfo {author} {\bibfnamefont {K.~J.}\
  \bibnamefont {Satzinger}}, \bibinfo {author} {\bibfnamefont {V.}~\bibnamefont
  {Smelyanskiy}}, \bibinfo {author} {\bibfnamefont {K.~J.}\ \bibnamefont
  {Sung}}, \bibinfo {author} {\bibfnamefont {M.~D.}\ \bibnamefont
  {Trevithick}}, \bibinfo {author} {\bibfnamefont {A.}~\bibnamefont
  {Vainsencher}}, \bibinfo {author} {\bibfnamefont {B.}~\bibnamefont
  {Villalonga}}, \bibinfo {author} {\bibfnamefont {T.}~\bibnamefont {White}},
  \bibinfo {author} {\bibfnamefont {Z.~J.}\ \bibnamefont {Yao}}, \bibinfo
  {author} {\bibfnamefont {P.}~\bibnamefont {Yeh}}, \bibinfo {author}
  {\bibfnamefont {A.}~\bibnamefont {Zalcman}}, \bibinfo {author} {\bibfnamefont
  {H.}~\bibnamefont {Neven}}, \ and\ \bibinfo {author} {\bibfnamefont {J.~M.}\
  \bibnamefont {Martinis}},\ }\href {\doibase 10.1038/s41586-019-1666-5}
  {\bibfield  {journal} {\bibinfo  {journal} {Nature}\ }\textbf {\bibinfo
  {volume} {574}},\ \bibinfo {pages} {505} (\bibinfo {year}
  {2019})}\BibitemShut {NoStop}%
\bibitem [{\citenamefont {Li}\ and\ \citenamefont
  {Benjamin}(2017)}]{li_efficient_2017}%
  \BibitemOpen
  \bibfield  {author} {\bibinfo {author} {\bibfnamefont {Y.}~\bibnamefont
  {Li}}\ and\ \bibinfo {author} {\bibfnamefont {S.~C.}\ \bibnamefont
  {Benjamin}},\ }\href {\doibase 10.1103/PhysRevX.7.021050} {\bibfield
  {journal} {\bibinfo  {journal} {Phys. Rev. X}\ }\textbf {\bibinfo {volume}
  {7}},\ \bibinfo {pages} {021050} (\bibinfo {year} {2017})}\BibitemShut
  {NoStop}%
\bibitem [{\citenamefont {Temme}\ \emph {et~al.}(2017)\citenamefont {Temme},
  \citenamefont {Bravyi},\ and\ \citenamefont {Gambetta}}]{temme_error_2017}%
  \BibitemOpen
  \bibfield  {author} {\bibinfo {author} {\bibfnamefont {K.}~\bibnamefont
  {Temme}}, \bibinfo {author} {\bibfnamefont {S.}~\bibnamefont {Bravyi}}, \
  and\ \bibinfo {author} {\bibfnamefont {J.~M.}\ \bibnamefont {Gambetta}},\
  }\href {\doibase 10.1103/PhysRevLett.119.180509} {\bibfield  {journal}
  {\bibinfo  {journal} {Phys. Rev. Lett.}\ }\textbf {\bibinfo {volume} {119}},\
  \bibinfo {pages} {180509} (\bibinfo {year} {2017})}\BibitemShut {NoStop}%
\bibitem [{\citenamefont {Abrams}\ and\ \citenamefont
  {Lloyd}(1997)}]{abrams_simulation_1997}%
  \BibitemOpen
  \bibfield  {author} {\bibinfo {author} {\bibfnamefont {D.~S.}\ \bibnamefont
  {Abrams}}\ and\ \bibinfo {author} {\bibfnamefont {S.}~\bibnamefont {Lloyd}},\
  }\href {\doibase 10.1103/PhysRevLett.79.2586} {\bibfield  {journal} {\bibinfo
   {journal} {Phys. Rev. Lett.}\ }\textbf {\bibinfo {volume} {79}},\ \bibinfo
  {pages} {2586} (\bibinfo {year} {1997})}\BibitemShut {NoStop}%
\bibitem [{\citenamefont {Abrams}\ and\ \citenamefont
  {Lloyd}(1999)}]{abrams_quantum_1999}%
  \BibitemOpen
  \bibfield  {author} {\bibinfo {author} {\bibfnamefont {D.~S.}\ \bibnamefont
  {Abrams}}\ and\ \bibinfo {author} {\bibfnamefont {S.}~\bibnamefont {Lloyd}},\
  }\href {\doibase 10.1103/PhysRevLett.83.5162} {\bibfield  {journal} {\bibinfo
   {journal} {Phys. Rev. Lett.}\ }\textbf {\bibinfo {volume} {83}},\ \bibinfo
  {pages} {5162} (\bibinfo {year} {1999})}\BibitemShut {NoStop}%
\bibitem [{\citenamefont {Peruzzo}\ \emph {et~al.}(2014)\citenamefont
  {Peruzzo}, \citenamefont {McClean}, \citenamefont {Shadbolt}, \citenamefont
  {Yung}, \citenamefont {Zhou}, \citenamefont {Love}, \citenamefont
  {Aspuru-Guzik},\ and\ \citenamefont {O’Brien}}]{peruzzo_variational_2014}%
  \BibitemOpen
  \bibfield  {author} {\bibinfo {author} {\bibfnamefont {A.}~\bibnamefont
  {Peruzzo}}, \bibinfo {author} {\bibfnamefont {J.}~\bibnamefont {McClean}},
  \bibinfo {author} {\bibfnamefont {P.}~\bibnamefont {Shadbolt}}, \bibinfo
  {author} {\bibfnamefont {M.-H.}\ \bibnamefont {Yung}}, \bibinfo {author}
  {\bibfnamefont {X.-Q.}\ \bibnamefont {Zhou}}, \bibinfo {author}
  {\bibfnamefont {P.~J.}\ \bibnamefont {Love}}, \bibinfo {author}
  {\bibfnamefont {A.}~\bibnamefont {Aspuru-Guzik}}, \ and\ \bibinfo {author}
  {\bibfnamefont {J.~L.}\ \bibnamefont {O’Brien}},\ }\href {\doibase
  10.1038/ncomms5213} {\bibfield  {journal} {\bibinfo  {journal} {Nature
  Communications}\ }\textbf {\bibinfo {volume} {5}},\ \bibinfo {pages} {1}
  (\bibinfo {year} {2014})}\BibitemShut {NoStop}%
\bibitem [{\citenamefont {McClean}\ \emph {et~al.}(2016)\citenamefont
  {McClean}, \citenamefont {Romero}, \citenamefont {Babbush},\ and\
  \citenamefont {Aspuru-Guzik}}]{mcclean_theory_2016}%
  \BibitemOpen
  \bibfield  {author} {\bibinfo {author} {\bibfnamefont {J.~R.}\ \bibnamefont
  {McClean}}, \bibinfo {author} {\bibfnamefont {J.}~\bibnamefont {Romero}},
  \bibinfo {author} {\bibfnamefont {R.}~\bibnamefont {Babbush}}, \ and\
  \bibinfo {author} {\bibfnamefont {A.}~\bibnamefont {Aspuru-Guzik}},\ }\href
  {\doibase 10.1088/1367-2630/18/2/023023} {\bibfield  {journal} {\bibinfo
  {journal} {New J. Phys.}\ }\textbf {\bibinfo {volume} {18}},\ \bibinfo
  {pages} {023023} (\bibinfo {year} {2016})},\ \bibinfo {note} {arXiv:
  1509.04279}\BibitemShut {NoStop}%
\bibitem [{\citenamefont {Dallaire-Demers}\ \emph {et~al.}(2019)\citenamefont
  {Dallaire-Demers}, \citenamefont {Romero}, \citenamefont {Veis},
  \citenamefont {Sim},\ and\ \citenamefont
  {Aspuru-Guzik}}]{dallaire-demers_low-depth_2019}%
  \BibitemOpen
  \bibfield  {author} {\bibinfo {author} {\bibfnamefont {P.-L.}\ \bibnamefont
  {Dallaire-Demers}}, \bibinfo {author} {\bibfnamefont {J.}~\bibnamefont
  {Romero}}, \bibinfo {author} {\bibfnamefont {L.}~\bibnamefont {Veis}},
  \bibinfo {author} {\bibfnamefont {S.}~\bibnamefont {Sim}}, \ and\ \bibinfo
  {author} {\bibfnamefont {A.}~\bibnamefont {Aspuru-Guzik}},\ }\href {\doibase
  10.1088/2058-9565/ab3951} {\bibfield  {journal} {\bibinfo  {journal} {Quantum
  Sci. Technol.}\ }\textbf {\bibinfo {volume} {4}},\ \bibinfo {pages} {045005}
  (\bibinfo {year} {2019})}\BibitemShut {NoStop}%
\bibitem [{\citenamefont {Barron}\ \emph {et~al.}(2020)\citenamefont {Barron},
  \citenamefont {Gard}, \citenamefont {Altman}, \citenamefont {Mayhall},
  \citenamefont {Barnes},\ and\ \citenamefont
  {Economou}}]{barron_preserving_2020}%
  \BibitemOpen
  \bibfield  {author} {\bibinfo {author} {\bibfnamefont {G.~S.}\ \bibnamefont
  {Barron}}, \bibinfo {author} {\bibfnamefont {B.~T.}\ \bibnamefont {Gard}},
  \bibinfo {author} {\bibfnamefont {O.~J.}\ \bibnamefont {Altman}}, \bibinfo
  {author} {\bibfnamefont {N.~J.}\ \bibnamefont {Mayhall}}, \bibinfo {author}
  {\bibfnamefont {E.}~\bibnamefont {Barnes}}, \ and\ \bibinfo {author}
  {\bibfnamefont {S.~E.}\ \bibnamefont {Economou}},\ }\href
  {http://arxiv.org/abs/2003.00171} {\bibfield  {journal} {\bibinfo  {journal}
  {arXiv:2003.00171 [quant-ph]}\ } (\bibinfo {year} {2020})},\ \bibinfo {note}
  {arXiv: 2003.00171}\BibitemShut {NoStop}%
\bibitem [{\citenamefont {Gard}\ \emph {et~al.}(2020)\citenamefont {Gard},
  \citenamefont {Zhu}, \citenamefont {Barron}, \citenamefont {Mayhall},
  \citenamefont {Economou},\ and\ \citenamefont
  {Barnes}}]{gard_efficient_2020}%
  \BibitemOpen
  \bibfield  {author} {\bibinfo {author} {\bibfnamefont {B.~T.}\ \bibnamefont
  {Gard}}, \bibinfo {author} {\bibfnamefont {L.}~\bibnamefont {Zhu}}, \bibinfo
  {author} {\bibfnamefont {G.~S.}\ \bibnamefont {Barron}}, \bibinfo {author}
  {\bibfnamefont {N.~J.}\ \bibnamefont {Mayhall}}, \bibinfo {author}
  {\bibfnamefont {S.~E.}\ \bibnamefont {Economou}}, \ and\ \bibinfo {author}
  {\bibfnamefont {E.}~\bibnamefont {Barnes}},\ }\href {\doibase
  10.1038/s41534-019-0240-1} {\bibfield  {journal} {\bibinfo  {journal} {npj
  Quantum Information}\ }\textbf {\bibinfo {volume} {6}},\ \bibinfo {pages} {1}
  (\bibinfo {year} {2020})}\BibitemShut {NoStop}%
\bibitem [{\citenamefont {Bartlett}\ \emph {et~al.}(1989)\citenamefont
  {Bartlett}, \citenamefont {Kucharski},\ and\ \citenamefont
  {Noga}}]{bartlett_alternative_1989}%
  \BibitemOpen
  \bibfield  {author} {\bibinfo {author} {\bibfnamefont {R.~J.}\ \bibnamefont
  {Bartlett}}, \bibinfo {author} {\bibfnamefont {S.~A.}\ \bibnamefont
  {Kucharski}}, \ and\ \bibinfo {author} {\bibfnamefont {J.}~\bibnamefont
  {Noga}},\ }\href {\doibase 10.1016/S0009-2614(89)87372-5} {\bibfield
  {journal} {\bibinfo  {journal} {Chemical Physics Letters}\ }\textbf {\bibinfo
  {volume} {155}},\ \bibinfo {pages} {133} (\bibinfo {year}
  {1989})}\BibitemShut {NoStop}%
\bibitem [{\citenamefont {Kutzelnigg}(1991)}]{kutzelnigg_error_1991}%
  \BibitemOpen
  \bibfield  {author} {\bibinfo {author} {\bibfnamefont {W.}~\bibnamefont
  {Kutzelnigg}},\ }\href {\doibase 10.1007/BF01117418} {\bibfield  {journal}
  {\bibinfo  {journal} {Theoret. Chim. Acta}\ }\textbf {\bibinfo {volume}
  {80}},\ \bibinfo {pages} {349} (\bibinfo {year} {1991})}\BibitemShut
  {NoStop}%
\bibitem [{\citenamefont {Taube}\ and\ \citenamefont
  {Bartlett}(2006)}]{taube_new_2006}%
  \BibitemOpen
  \bibfield  {author} {\bibinfo {author} {\bibfnamefont {A.~G.}\ \bibnamefont
  {Taube}}\ and\ \bibinfo {author} {\bibfnamefont {R.~J.}\ \bibnamefont
  {Bartlett}},\ }\href {\doibase 10.1002/qua.21198} {\bibfield  {journal}
  {\bibinfo  {journal} {International Journal of Quantum Chemistry}\ }\textbf
  {\bibinfo {volume} {106}},\ \bibinfo {pages} {3393} (\bibinfo {year}
  {2006})}\BibitemShut {NoStop}%
\bibitem [{\citenamefont {Cooper}\ and\ \citenamefont
  {Knowles}(2010)}]{cooper_benchmark_2010}%
  \BibitemOpen
  \bibfield  {author} {\bibinfo {author} {\bibfnamefont {B.}~\bibnamefont
  {Cooper}}\ and\ \bibinfo {author} {\bibfnamefont {P.~J.}\ \bibnamefont
  {Knowles}},\ }\href {\doibase 10.1063/1.3520564} {\bibfield  {journal}
  {\bibinfo  {journal} {J. Chem. Phys.}\ }\textbf {\bibinfo {volume} {133}},\
  \bibinfo {pages} {234102} (\bibinfo {year} {2010})}\BibitemShut {NoStop}%
\bibitem [{\citenamefont {Evangelista}(2011)}]{evangelista_alternative_2011}%
  \BibitemOpen
  \bibfield  {author} {\bibinfo {author} {\bibfnamefont {F.~A.}\ \bibnamefont
  {Evangelista}},\ }\href {\doibase 10.1063/1.3598471} {\bibfield  {journal}
  {\bibinfo  {journal} {J. Chem. Phys.}\ }\textbf {\bibinfo {volume} {134}},\
  \bibinfo {pages} {224102} (\bibinfo {year} {2011})}\BibitemShut {NoStop}%
\bibitem [{\citenamefont {Barkoutsos}\ \emph {et~al.}(2018)\citenamefont
  {Barkoutsos}, \citenamefont {Gonthier}, \citenamefont {Sokolov},
  \citenamefont {Moll}, \citenamefont {Salis}, \citenamefont {Fuhrer},
  \citenamefont {Ganzhorn}, \citenamefont {Egger}, \citenamefont {Troyer},
  \citenamefont {Mezzacapo}, \citenamefont {Filipp},\ and\ \citenamefont
  {Tavernelli}}]{barkoutsos_quantum_2018}%
  \BibitemOpen
  \bibfield  {author} {\bibinfo {author} {\bibfnamefont {P.~K.}\ \bibnamefont
  {Barkoutsos}}, \bibinfo {author} {\bibfnamefont {J.~F.}\ \bibnamefont
  {Gonthier}}, \bibinfo {author} {\bibfnamefont {I.}~\bibnamefont {Sokolov}},
  \bibinfo {author} {\bibfnamefont {N.}~\bibnamefont {Moll}}, \bibinfo {author}
  {\bibfnamefont {G.}~\bibnamefont {Salis}}, \bibinfo {author} {\bibfnamefont
  {A.}~\bibnamefont {Fuhrer}}, \bibinfo {author} {\bibfnamefont
  {M.}~\bibnamefont {Ganzhorn}}, \bibinfo {author} {\bibfnamefont {D.~J.}\
  \bibnamefont {Egger}}, \bibinfo {author} {\bibfnamefont {M.}~\bibnamefont
  {Troyer}}, \bibinfo {author} {\bibfnamefont {A.}~\bibnamefont {Mezzacapo}},
  \bibinfo {author} {\bibfnamefont {S.}~\bibnamefont {Filipp}}, \ and\ \bibinfo
  {author} {\bibfnamefont {I.}~\bibnamefont {Tavernelli}},\ }\href {\doibase
  10.1103/PhysRevA.98.022322} {\bibfield  {journal} {\bibinfo  {journal} {Phys.
  Rev. A}\ }\textbf {\bibinfo {volume} {98}},\ \bibinfo {pages} {022322}
  (\bibinfo {year} {2018})}\BibitemShut {NoStop}%
\bibitem [{\citenamefont {Romero}\ \emph {et~al.}(2018)\citenamefont {Romero},
  \citenamefont {Babbush}, \citenamefont {McClean}, \citenamefont {Hempel},
  \citenamefont {Love},\ and\ \citenamefont
  {Aspuru-Guzik}}]{romero_strategies_2018}%
  \BibitemOpen
  \bibfield  {author} {\bibinfo {author} {\bibfnamefont {J.}~\bibnamefont
  {Romero}}, \bibinfo {author} {\bibfnamefont {R.}~\bibnamefont {Babbush}},
  \bibinfo {author} {\bibfnamefont {J.~R.}\ \bibnamefont {McClean}}, \bibinfo
  {author} {\bibfnamefont {C.}~\bibnamefont {Hempel}}, \bibinfo {author}
  {\bibfnamefont {P.~J.}\ \bibnamefont {Love}}, \ and\ \bibinfo {author}
  {\bibfnamefont {A.}~\bibnamefont {Aspuru-Guzik}},\ }\href {\doibase
  10.1088/2058-9565/aad3e4} {\bibfield  {journal} {\bibinfo  {journal} {Quantum
  Sci. Technol.}\ }\textbf {\bibinfo {volume} {4}},\ \bibinfo {pages} {014008}
  (\bibinfo {year} {2018})}\BibitemShut {NoStop}%
\bibitem [{\citenamefont {Harsha}\ \emph {et~al.}(2018)\citenamefont {Harsha},
  \citenamefont {Shiozaki},\ and\ \citenamefont
  {Scuseria}}]{harsha_difference_2018}%
  \BibitemOpen
  \bibfield  {author} {\bibinfo {author} {\bibfnamefont {G.}~\bibnamefont
  {Harsha}}, \bibinfo {author} {\bibfnamefont {T.}~\bibnamefont {Shiozaki}}, \
  and\ \bibinfo {author} {\bibfnamefont {G.~E.}\ \bibnamefont {Scuseria}},\
  }\href {\doibase 10.1063/1.5011033} {\bibfield  {journal} {\bibinfo
  {journal} {J. Chem. Phys.}\ }\textbf {\bibinfo {volume} {148}},\ \bibinfo
  {pages} {044107} (\bibinfo {year} {2018})}\BibitemShut {NoStop}%
\bibitem [{\citenamefont {Grimsley}\ \emph {et~al.}(2019)\citenamefont
  {Grimsley}, \citenamefont {Economou}, \citenamefont {Barnes},\ and\
  \citenamefont {Mayhall}}]{grimsley_adaptive_2019}%
  \BibitemOpen
  \bibfield  {author} {\bibinfo {author} {\bibfnamefont {H.~R.}\ \bibnamefont
  {Grimsley}}, \bibinfo {author} {\bibfnamefont {S.~E.}\ \bibnamefont
  {Economou}}, \bibinfo {author} {\bibfnamefont {E.}~\bibnamefont {Barnes}}, \
  and\ \bibinfo {author} {\bibfnamefont {N.~J.}\ \bibnamefont {Mayhall}},\
  }\href {\doibase 10.1038/s41467-019-10988-2} {\bibfield  {journal} {\bibinfo
  {journal} {Nat Commun}\ }\textbf {\bibinfo {volume} {10}},\ \bibinfo {pages}
  {3007} (\bibinfo {year} {2019})}\BibitemShut {NoStop}%
\bibitem [{\citenamefont {Lee}\ \emph {et~al.}(2019)\citenamefont {Lee},
  \citenamefont {Huggins}, \citenamefont {Head-Gordon},\ and\ \citenamefont
  {Whaley}}]{lee_generalized_2019}%
  \BibitemOpen
  \bibfield  {author} {\bibinfo {author} {\bibfnamefont {J.}~\bibnamefont
  {Lee}}, \bibinfo {author} {\bibfnamefont {W.~J.}\ \bibnamefont {Huggins}},
  \bibinfo {author} {\bibfnamefont {M.}~\bibnamefont {Head-Gordon}}, \ and\
  \bibinfo {author} {\bibfnamefont {K.~B.}\ \bibnamefont {Whaley}},\ }\href
  {\doibase 10.1021/acs.jctc.8b01004} {\bibfield  {journal} {\bibinfo
  {journal} {J. Chem. Theory Comput.}\ }\textbf {\bibinfo {volume} {15}},\
  \bibinfo {pages} {311} (\bibinfo {year} {2019})}\BibitemShut {NoStop}%
\bibitem [{\citenamefont {Kandala}\ \emph {et~al.}(2017)\citenamefont
  {Kandala}, \citenamefont {Mezzacapo}, \citenamefont {Temme}, \citenamefont
  {Takita}, \citenamefont {Brink}, \citenamefont {Chow},\ and\ \citenamefont
  {Gambetta}}]{kandala_hardware-efficient_2017}%
  \BibitemOpen
  \bibfield  {author} {\bibinfo {author} {\bibfnamefont {A.}~\bibnamefont
  {Kandala}}, \bibinfo {author} {\bibfnamefont {A.}~\bibnamefont {Mezzacapo}},
  \bibinfo {author} {\bibfnamefont {K.}~\bibnamefont {Temme}}, \bibinfo
  {author} {\bibfnamefont {M.}~\bibnamefont {Takita}}, \bibinfo {author}
  {\bibfnamefont {M.}~\bibnamefont {Brink}}, \bibinfo {author} {\bibfnamefont
  {J.~M.}\ \bibnamefont {Chow}}, \ and\ \bibinfo {author} {\bibfnamefont
  {J.~M.}\ \bibnamefont {Gambetta}},\ }\href {\doibase 10.1038/nature23879}
  {\bibfield  {journal} {\bibinfo  {journal} {Nature}\ }\textbf {\bibinfo
  {volume} {549}},\ \bibinfo {pages} {242} (\bibinfo {year}
  {2017})}\BibitemShut {NoStop}%
\bibitem [{\citenamefont {McClean}\ \emph {et~al.}(2018)\citenamefont
  {McClean}, \citenamefont {Boixo}, \citenamefont {Smelyanskiy}, \citenamefont
  {Babbush},\ and\ \citenamefont {Neven}}]{mcclean_barren_2018}%
  \BibitemOpen
  \bibfield  {author} {\bibinfo {author} {\bibfnamefont {J.~R.}\ \bibnamefont
  {McClean}}, \bibinfo {author} {\bibfnamefont {S.}~\bibnamefont {Boixo}},
  \bibinfo {author} {\bibfnamefont {V.~N.}\ \bibnamefont {Smelyanskiy}},
  \bibinfo {author} {\bibfnamefont {R.}~\bibnamefont {Babbush}}, \ and\
  \bibinfo {author} {\bibfnamefont {H.}~\bibnamefont {Neven}},\ }\href
  {\doibase 10.1038/s41467-018-07090-4} {\bibfield  {journal} {\bibinfo
  {journal} {Nature Communications}\ }\textbf {\bibinfo {volume} {9}},\
  \bibinfo {pages} {1} (\bibinfo {year} {2018})}\BibitemShut {NoStop}%
\bibitem [{\citenamefont {Grimsley}\ \emph {et~al.}(2020)\citenamefont
  {Grimsley}, \citenamefont {Claudino}, \citenamefont {Economou}, \citenamefont
  {Barnes},\ and\ \citenamefont {Mayhall}}]{grimsley_is_2020}%
  \BibitemOpen
  \bibfield  {author} {\bibinfo {author} {\bibfnamefont {H.~R.}\ \bibnamefont
  {Grimsley}}, \bibinfo {author} {\bibfnamefont {D.}~\bibnamefont {Claudino}},
  \bibinfo {author} {\bibfnamefont {S.~E.}\ \bibnamefont {Economou}}, \bibinfo
  {author} {\bibfnamefont {E.}~\bibnamefont {Barnes}}, \ and\ \bibinfo {author}
  {\bibfnamefont {N.~J.}\ \bibnamefont {Mayhall}},\ }\href {\doibase
  10.1021/acs.jctc.9b01083} {\bibfield  {journal} {\bibinfo  {journal} {J.
  Chem. Theory Comput.}\ }\textbf {\bibinfo {volume} {16}},\ \bibinfo {pages}
  {1} (\bibinfo {year} {2020})}\BibitemShut {NoStop}%
\bibitem [{\citenamefont {Tang}\ \emph {et~al.}(2020)\citenamefont {Tang},
  \citenamefont {Shkolnikov}, \citenamefont {Barron}, \citenamefont {Grimsley},
  \citenamefont {Mayhall}, \citenamefont {Barnes},\ and\ \citenamefont
  {Economou}}]{tang_qubit-adapt-vqe_2020}%
  \BibitemOpen
  \bibfield  {author} {\bibinfo {author} {\bibfnamefont {H.~L.}\ \bibnamefont
  {Tang}}, \bibinfo {author} {\bibfnamefont {V.~O.}\ \bibnamefont
  {Shkolnikov}}, \bibinfo {author} {\bibfnamefont {G.~S.}\ \bibnamefont
  {Barron}}, \bibinfo {author} {\bibfnamefont {H.~R.}\ \bibnamefont
  {Grimsley}}, \bibinfo {author} {\bibfnamefont {N.~J.}\ \bibnamefont
  {Mayhall}}, \bibinfo {author} {\bibfnamefont {E.}~\bibnamefont {Barnes}}, \
  and\ \bibinfo {author} {\bibfnamefont {S.~E.}\ \bibnamefont {Economou}},\
  }\href {http://arxiv.org/abs/1911.10205} {\bibfield  {journal} {\bibinfo
  {journal} {arXiv:1911.10205 [quant-ph]}\ } (\bibinfo {year} {2020})},\
  \bibinfo {note} {arXiv: 1911.10205}\BibitemShut {NoStop}%
\bibitem [{\citenamefont {Jiménez-Hoyos}\ \emph {et~al.}(2012)\citenamefont
  {Jiménez-Hoyos}, \citenamefont {Henderson}, \citenamefont {Tsuchimochi},\
  and\ \citenamefont {Scuseria}}]{jimenez-hoyos_projected_2012}%
  \BibitemOpen
  \bibfield  {author} {\bibinfo {author} {\bibfnamefont {C.~A.}\ \bibnamefont
  {Jiménez-Hoyos}}, \bibinfo {author} {\bibfnamefont {T.~M.}\ \bibnamefont
  {Henderson}}, \bibinfo {author} {\bibfnamefont {T.}~\bibnamefont
  {Tsuchimochi}}, \ and\ \bibinfo {author} {\bibfnamefont {G.~E.}\ \bibnamefont
  {Scuseria}},\ }\href {\doibase 10.1063/1.4705280} {\bibfield  {journal}
  {\bibinfo  {journal} {J. Chem. Phys.}\ }\textbf {\bibinfo {volume} {136}},\
  \bibinfo {pages} {164109} (\bibinfo {year} {2012})}\BibitemShut {NoStop}%
\bibitem [{\citenamefont {Bulik}\ \emph {et~al.}(2015)\citenamefont {Bulik},
  \citenamefont {Henderson},\ and\ \citenamefont {Scuseria}}]{bulik_can_2015}%
  \BibitemOpen
  \bibfield  {author} {\bibinfo {author} {\bibfnamefont {I.~W.}\ \bibnamefont
  {Bulik}}, \bibinfo {author} {\bibfnamefont {T.~M.}\ \bibnamefont
  {Henderson}}, \ and\ \bibinfo {author} {\bibfnamefont {G.~E.}\ \bibnamefont
  {Scuseria}},\ }\href {\doibase 10.1021/acs.jctc.5b00422} {\bibfield
  {journal} {\bibinfo  {journal} {J. Chem. Theory Comput.}\ }\textbf {\bibinfo
  {volume} {11}},\ \bibinfo {pages} {3171} (\bibinfo {year}
  {2015})}\BibitemShut {NoStop}%
\bibitem [{\citenamefont {Tsuchimochi}\ \emph {et~al.}(2020)\citenamefont
  {Tsuchimochi}, \citenamefont {Mori},\ and\ \citenamefont
  {Ten-no}}]{tsuchimochi_exact_2020}%
  \BibitemOpen
  \bibfield  {author} {\bibinfo {author} {\bibfnamefont {T.}~\bibnamefont
  {Tsuchimochi}}, \bibinfo {author} {\bibfnamefont {Y.}~\bibnamefont {Mori}}, \
  and\ \bibinfo {author} {\bibfnamefont {S.~L.}\ \bibnamefont {Ten-no}},\
  }\href {http://arxiv.org/abs/2004.12024} {\bibfield  {journal} {\bibinfo
  {journal} {arXiv:2004.12024 [cond-mat, physics:physics, physics:quant-ph]}\ }
  (\bibinfo {year} {2020})},\ \bibinfo {note} {arXiv: 2004.12024}\BibitemShut
  {NoStop}%
\bibitem [{\citenamefont {Lacroix}(2020)}]{lacroix_symmetry_2020}%
  \BibitemOpen
  \bibfield  {author} {\bibinfo {author} {\bibfnamefont {D.}~\bibnamefont
  {Lacroix}},\ }\href {http://arxiv.org/abs/2006.06491} {\bibfield  {journal}
  {\bibinfo  {journal} {arXiv:2006.06491 [nucl-th, physics:quant-ph]}\ }
  (\bibinfo {year} {2020})},\ \bibinfo {note} {arXiv: 2006.06491}\BibitemShut
  {NoStop}%
\bibitem [{\citenamefont {Bardeen}\ \emph {et~al.}(1957)\citenamefont
  {Bardeen}, \citenamefont {Cooper},\ and\ \citenamefont
  {Schrieffer}}]{bardeen_theory_1957}%
  \BibitemOpen
  \bibfield  {author} {\bibinfo {author} {\bibfnamefont {J.}~\bibnamefont
  {Bardeen}}, \bibinfo {author} {\bibfnamefont {L.~N.}\ \bibnamefont {Cooper}},
  \ and\ \bibinfo {author} {\bibfnamefont {J.~R.}\ \bibnamefont {Schrieffer}},\
  }\href {\doibase 10.1103/PhysRev.108.1175} {\bibfield  {journal} {\bibinfo
  {journal} {Phys. Rev.}\ }\textbf {\bibinfo {volume} {108}},\ \bibinfo {pages}
  {1175} (\bibinfo {year} {1957})}\BibitemShut {NoStop}%
\bibitem [{\citenamefont {Bayman}(1960)}]{bayman_derivation_1960}%
  \BibitemOpen
  \bibfield  {author} {\bibinfo {author} {\bibfnamefont {B.~F.}\ \bibnamefont
  {Bayman}},\ }\href {\doibase 10.1016/0029-5582(60)90279-0} {\bibfield
  {journal} {\bibinfo  {journal} {Nuclear Physics}\ }\textbf {\bibinfo {volume}
  {15}},\ \bibinfo {pages} {33} (\bibinfo {year} {1960})}\BibitemShut {NoStop}%
\bibitem [{\citenamefont {Sierra}\ \emph {et~al.}(2000)\citenamefont {Sierra},
  \citenamefont {Dukelsky}, \citenamefont {Dussel}, \citenamefont {von Delft},\
  and\ \citenamefont {Braun}}]{sierra_exact_2000}%
  \BibitemOpen
  \bibfield  {author} {\bibinfo {author} {\bibfnamefont {G.}~\bibnamefont
  {Sierra}}, \bibinfo {author} {\bibfnamefont {J.}~\bibnamefont {Dukelsky}},
  \bibinfo {author} {\bibfnamefont {G.~G.}\ \bibnamefont {Dussel}}, \bibinfo
  {author} {\bibfnamefont {J.}~\bibnamefont {von Delft}}, \ and\ \bibinfo
  {author} {\bibfnamefont {F.}~\bibnamefont {Braun}},\ }\href {\doibase
  10.1103/PhysRevB.61.R11890} {\bibfield  {journal} {\bibinfo  {journal} {Phys.
  Rev. B}\ }\textbf {\bibinfo {volume} {61}},\ \bibinfo {pages} {R11890}
  (\bibinfo {year} {2000})}\BibitemShut {NoStop}%
\bibitem [{\citenamefont {Dukelsky}\ \emph {et~al.}(2004)\citenamefont
  {Dukelsky}, \citenamefont {Pittel},\ and\ \citenamefont
  {Sierra}}]{dukelsky_colloquium:_2004}%
  \BibitemOpen
  \bibfield  {author} {\bibinfo {author} {\bibfnamefont {J.}~\bibnamefont
  {Dukelsky}}, \bibinfo {author} {\bibfnamefont {S.}~\bibnamefont {Pittel}}, \
  and\ \bibinfo {author} {\bibfnamefont {G.}~\bibnamefont {Sierra}},\ }\href
  {\doibase 10.1103/RevModPhys.76.643} {\bibfield  {journal} {\bibinfo
  {journal} {Rev. Mod. Phys.}\ }\textbf {\bibinfo {volume} {76}},\ \bibinfo
  {pages} {643} (\bibinfo {year} {2004})}\BibitemShut {NoStop}%
\bibitem [{\citenamefont {Bytautas}\ \emph {et~al.}(2011)\citenamefont
  {Bytautas}, \citenamefont {Henderson}, \citenamefont {Jiménez-Hoyos},
  \citenamefont {Ellis},\ and\ \citenamefont
  {Scuseria}}]{bytautas_seniority_2011}%
  \BibitemOpen
  \bibfield  {author} {\bibinfo {author} {\bibfnamefont {L.}~\bibnamefont
  {Bytautas}}, \bibinfo {author} {\bibfnamefont {T.~M.}\ \bibnamefont
  {Henderson}}, \bibinfo {author} {\bibfnamefont {C.~A.}\ \bibnamefont
  {Jiménez-Hoyos}}, \bibinfo {author} {\bibfnamefont {J.~K.}\ \bibnamefont
  {Ellis}}, \ and\ \bibinfo {author} {\bibfnamefont {G.~E.}\ \bibnamefont
  {Scuseria}},\ }\href {\doibase 10.1063/1.3613706} {\bibfield  {journal}
  {\bibinfo  {journal} {J. Chem. Phys.}\ }\textbf {\bibinfo {volume} {135}},\
  \bibinfo {pages} {044119} (\bibinfo {year} {2011})}\BibitemShut {NoStop}%
\bibitem [{\citenamefont {Richardson}(1963)}]{richardson_restricted_1963}%
  \BibitemOpen
  \bibfield  {author} {\bibinfo {author} {\bibfnamefont {R.~W.}\ \bibnamefont
  {Richardson}},\ }\href {\doibase 10.1016/0031-9163(63)90259-2} {\bibfield
  {journal} {\bibinfo  {journal} {Physics Letters}\ }\textbf {\bibinfo {volume}
  {3}},\ \bibinfo {pages} {277} (\bibinfo {year} {1963})}\BibitemShut {NoStop}%
\bibitem [{\citenamefont {Henderson}\ \emph {et~al.}(2014)\citenamefont
  {Henderson}, \citenamefont {Scuseria}, \citenamefont {Dukelsky},
  \citenamefont {Signoracci},\ and\ \citenamefont
  {Duguet}}]{henderson_quasiparticle_2014}%
  \BibitemOpen
  \bibfield  {author} {\bibinfo {author} {\bibfnamefont {T.~M.}\ \bibnamefont
  {Henderson}}, \bibinfo {author} {\bibfnamefont {G.~E.}\ \bibnamefont
  {Scuseria}}, \bibinfo {author} {\bibfnamefont {J.}~\bibnamefont {Dukelsky}},
  \bibinfo {author} {\bibfnamefont {A.}~\bibnamefont {Signoracci}}, \ and\
  \bibinfo {author} {\bibfnamefont {T.}~\bibnamefont {Duguet}},\ }\href
  {\doibase 10.1103/PhysRevC.89.054305} {\bibfield  {journal} {\bibinfo
  {journal} {Phys. Rev. C}\ }\textbf {\bibinfo {volume} {89}},\ \bibinfo
  {pages} {054305} (\bibinfo {year} {2014})}\BibitemShut {NoStop}%
\bibitem [{\citenamefont {Henderson}\ \emph {et~al.}(2015)\citenamefont
  {Henderson}, \citenamefont {Bulik},\ and\ \citenamefont
  {Scuseria}}]{henderson_pair_2015}%
  \BibitemOpen
  \bibfield  {author} {\bibinfo {author} {\bibfnamefont {T.~M.}\ \bibnamefont
  {Henderson}}, \bibinfo {author} {\bibfnamefont {I.~W.}\ \bibnamefont
  {Bulik}}, \ and\ \bibinfo {author} {\bibfnamefont {G.~E.}\ \bibnamefont
  {Scuseria}},\ }\href {\doibase 10.1063/1.4921986} {\bibfield  {journal}
  {\bibinfo  {journal} {J. Chem. Phys.}\ }\textbf {\bibinfo {volume} {142}},\
  \bibinfo {pages} {214116} (\bibinfo {year} {2015})}\BibitemShut {NoStop}%
\bibitem [{\citenamefont {Degroote}\ \emph {et~al.}(2016)\citenamefont
  {Degroote}, \citenamefont {Henderson}, \citenamefont {Zhao}, \citenamefont
  {Dukelsky},\ and\ \citenamefont {Scuseria}}]{degroote_polynomial_2016}%
  \BibitemOpen
  \bibfield  {author} {\bibinfo {author} {\bibfnamefont {M.}~\bibnamefont
  {Degroote}}, \bibinfo {author} {\bibfnamefont {T.~M.}\ \bibnamefont
  {Henderson}}, \bibinfo {author} {\bibfnamefont {J.}~\bibnamefont {Zhao}},
  \bibinfo {author} {\bibfnamefont {J.}~\bibnamefont {Dukelsky}}, \ and\
  \bibinfo {author} {\bibfnamefont {G.~E.}\ \bibnamefont {Scuseria}},\ }\href
  {\doibase 10.1103/PhysRevB.93.125124} {\bibfield  {journal} {\bibinfo
  {journal} {Phys. Rev. B}\ }\textbf {\bibinfo {volume} {93}},\ \bibinfo
  {pages} {125124} (\bibinfo {year} {2016})}\BibitemShut {NoStop}%
\bibitem [{\citenamefont {Qiu}\ \emph {et~al.}(2019)\citenamefont {Qiu},
  \citenamefont {Henderson}, \citenamefont {Duguet},\ and\ \citenamefont
  {Scuseria}}]{qiu_particle-number_2019}%
  \BibitemOpen
  \bibfield  {author} {\bibinfo {author} {\bibfnamefont {Y.}~\bibnamefont
  {Qiu}}, \bibinfo {author} {\bibfnamefont {T.~M.}\ \bibnamefont {Henderson}},
  \bibinfo {author} {\bibfnamefont {T.}~\bibnamefont {Duguet}}, \ and\ \bibinfo
  {author} {\bibfnamefont {G.~E.}\ \bibnamefont {Scuseria}},\ }\href {\doibase
  10.1103/PhysRevC.99.044301} {\bibfield  {journal} {\bibinfo  {journal} {Phys.
  Rev. C}\ }\textbf {\bibinfo {volume} {99}},\ \bibinfo {pages} {044301}
  (\bibinfo {year} {2019})}\BibitemShut {NoStop}%
\bibitem [{\citenamefont {Henderson}\ and\ \citenamefont
  {Scuseria}(2020)}]{henderson_correlating_2020}%
  \BibitemOpen
  \bibfield  {author} {\bibinfo {author} {\bibfnamefont {T.~M.}\ \bibnamefont
  {Henderson}}\ and\ \bibinfo {author} {\bibfnamefont {G.~E.}\ \bibnamefont
  {Scuseria}},\ }\href {http://arxiv.org/abs/2007.03671} {\bibfield  {journal}
  {\bibinfo  {journal} {arXiv:2007.03671 [cond-mat, physics:physics]}\ }
  (\bibinfo {year} {2020})},\ \bibinfo {note} {arXiv: 2007.03671}\BibitemShut
  {NoStop}%
\bibitem [{\citenamefont {Henderson}\ and\ \citenamefont
  {Scuseria}(2019)}]{henderson_geminal-based_2019}%
  \BibitemOpen
  \bibfield  {author} {\bibinfo {author} {\bibfnamefont {T.~M.}\ \bibnamefont
  {Henderson}}\ and\ \bibinfo {author} {\bibfnamefont {G.~E.}\ \bibnamefont
  {Scuseria}},\ }\href {\doibase 10.1063/1.5116715} {\bibfield  {journal}
  {\bibinfo  {journal} {J. Chem. Phys.}\ }\textbf {\bibinfo {volume} {151}},\
  \bibinfo {pages} {051101} (\bibinfo {year} {2019})}\BibitemShut {NoStop}%
\bibitem [{\citenamefont {Dutta}\ \emph {et~al.}(2020)\citenamefont {Dutta},
  \citenamefont {Henderson},\ and\ \citenamefont
  {Scuseria}}]{dutta_geminal_2020}%
  \BibitemOpen
  \bibfield  {author} {\bibinfo {author} {\bibfnamefont {R.}~\bibnamefont
  {Dutta}}, \bibinfo {author} {\bibfnamefont {T.~M.}\ \bibnamefont
  {Henderson}}, \ and\ \bibinfo {author} {\bibfnamefont {G.~E.}\ \bibnamefont
  {Scuseria}},\ }\href {http://arxiv.org/abs/2008.00552} {\bibfield  {journal}
  {\bibinfo  {journal} {arXiv:2008.00552 [physics]}\ } (\bibinfo {year}
  {2020})},\ \bibinfo {note} {arXiv: 2008.00552}\BibitemShut {NoStop}%
\bibitem [{\citenamefont {Ring}\ and\ \citenamefont
  {Schuck}(1980)}]{ring_nuclear_1980}%
  \BibitemOpen
  \bibfield  {author} {\bibinfo {author} {\bibfnamefont {P.}~\bibnamefont
  {Ring}}\ and\ \bibinfo {author} {\bibfnamefont {P.}~\bibnamefont {Schuck}},\
  }\href {https://www.springer.com/gp/book/9783540212065} {\emph {\bibinfo
  {title} {The {Nuclear} {Many}-{Body} {Problem}}}},\ Theoretical and
  {Mathematical} {Physics}, {The} {Nuclear} {Many}-{Body} {Problem}\ (\bibinfo
  {publisher} {Springer-Verlag},\ \bibinfo {address} {Berlin Heidelberg},\
  \bibinfo {year} {1980})\BibitemShut {NoStop}%
\bibitem [{\citenamefont {Blaizot}\ and\ \citenamefont
  {Ripka}(1986)}]{blaizot_quantum_1986}%
  \BibitemOpen
  \bibfield  {author} {\bibinfo {author} {\bibfnamefont {J.-P.}\ \bibnamefont
  {Blaizot}}\ and\ \bibinfo {author} {\bibfnamefont {G.}~\bibnamefont
  {Ripka}},\ }\href@noop {} {\emph {\bibinfo {title} {Quantum theory of finite
  systems}}}\ (\bibinfo  {publisher} {MIT Press},\ \bibinfo {address}
  {Cambridge, Mass.},\ \bibinfo {year} {1986})\ \bibinfo {note} {oCLC:
  11865956}\BibitemShut {NoStop}%
\bibitem [{\citenamefont {Dukelsky}\ \emph {et~al.}(2016)\citenamefont
  {Dukelsky}, \citenamefont {Pittel},\ and\ \citenamefont
  {Esebbag}}]{dukelsky_structure_2016}%
  \BibitemOpen
  \bibfield  {author} {\bibinfo {author} {\bibfnamefont {J.}~\bibnamefont
  {Dukelsky}}, \bibinfo {author} {\bibfnamefont {S.}~\bibnamefont {Pittel}}, \
  and\ \bibinfo {author} {\bibfnamefont {C.}~\bibnamefont {Esebbag}},\ }\href
  {\doibase 10.1103/PhysRevC.93.034313} {\bibfield  {journal} {\bibinfo
  {journal} {Phys. Rev. C}\ }\textbf {\bibinfo {volume} {93}},\ \bibinfo
  {pages} {034313} (\bibinfo {year} {2016})},\ \bibinfo {note} {arXiv:
  1603.00642}\BibitemShut {NoStop}%
\bibitem [{\citenamefont {Yang}(1962)}]{yang_concept_1962}%
  \BibitemOpen
  \bibfield  {author} {\bibinfo {author} {\bibfnamefont {C.~N.}\ \bibnamefont
  {Yang}},\ }\href {\doibase 10.1103/RevModPhys.34.694} {\bibfield  {journal}
  {\bibinfo  {journal} {Rev. Mod. Phys.}\ }\textbf {\bibinfo {volume} {34}},\
  \bibinfo {pages} {694} (\bibinfo {year} {1962})}\BibitemShut {NoStop}%
\bibitem [{\citenamefont {Veillard}\ and\ \citenamefont
  {Clementi}(1967)}]{veillard_complete_1967}%
  \BibitemOpen
  \bibfield  {author} {\bibinfo {author} {\bibfnamefont {A.}~\bibnamefont
  {Veillard}}\ and\ \bibinfo {author} {\bibfnamefont {E.}~\bibnamefont
  {Clementi}},\ }\href {\doibase 10.1007/BF01151915} {\bibfield  {journal}
  {\bibinfo  {journal} {Theoret. Chim. Acta}\ }\textbf {\bibinfo {volume}
  {7}},\ \bibinfo {pages} {133} (\bibinfo {year} {1967})}\BibitemShut {NoStop}%
\bibitem [{\citenamefont {Couty}\ and\ \citenamefont
  {Hall}(1997)}]{couty_generalized_1997}%
  \BibitemOpen
  \bibfield  {author} {\bibinfo {author} {\bibfnamefont {M.}~\bibnamefont
  {Couty}}\ and\ \bibinfo {author} {\bibfnamefont {M.~B.}\ \bibnamefont
  {Hall}},\ }\href {\doibase 10.1021/jp963953l} {\bibfield  {journal} {\bibinfo
   {journal} {J. Phys. Chem. A}\ }\textbf {\bibinfo {volume} {101}},\ \bibinfo
  {pages} {6936} (\bibinfo {year} {1997})}\BibitemShut {NoStop}%
\bibitem [{\citenamefont {Kollmar}\ and\ \citenamefont
  {Heß}(2003)}]{kollmar_new_2003}%
  \BibitemOpen
  \bibfield  {author} {\bibinfo {author} {\bibfnamefont {C.}~\bibnamefont
  {Kollmar}}\ and\ \bibinfo {author} {\bibfnamefont {B.~A.}\ \bibnamefont
  {Heß}},\ }\href {\doibase 10.1063/1.1590635} {\bibfield  {journal} {\bibinfo
   {journal} {J. Chem. Phys.}\ }\textbf {\bibinfo {volume} {119}},\ \bibinfo
  {pages} {4655} (\bibinfo {year} {2003})}\BibitemShut {NoStop}%
\bibitem [{\citenamefont {Sheikh}\ and\ \citenamefont
  {Ring}(2000)}]{sheikh_symmetry-projected_2000}%
  \BibitemOpen
  \bibfield  {author} {\bibinfo {author} {\bibfnamefont {J.~A.}\ \bibnamefont
  {Sheikh}}\ and\ \bibinfo {author} {\bibfnamefont {P.}~\bibnamefont {Ring}},\
  }\href {\doibase 10.1016/S0375-9474(99)00424-8} {\bibfield  {journal}
  {\bibinfo  {journal} {Nuclear Physics A}\ }\textbf {\bibinfo {volume}
  {665}},\ \bibinfo {pages} {71} (\bibinfo {year} {2000})}\BibitemShut
  {NoStop}%
\bibitem [{\citenamefont {Scuseria}\ \emph {et~al.}(2011)\citenamefont
  {Scuseria}, \citenamefont {Jiménez-Hoyos}, \citenamefont {Henderson},
  \citenamefont {Samanta},\ and\ \citenamefont
  {Ellis}}]{scuseria_projected_2011}%
  \BibitemOpen
  \bibfield  {author} {\bibinfo {author} {\bibfnamefont {G.~E.}\ \bibnamefont
  {Scuseria}}, \bibinfo {author} {\bibfnamefont {C.~A.}\ \bibnamefont
  {Jiménez-Hoyos}}, \bibinfo {author} {\bibfnamefont {T.~M.}\ \bibnamefont
  {Henderson}}, \bibinfo {author} {\bibfnamefont {K.}~\bibnamefont {Samanta}},
  \ and\ \bibinfo {author} {\bibfnamefont {J.~K.}\ \bibnamefont {Ellis}},\
  }\href {\doibase 10.1063/1.3643338} {\bibfield  {journal} {\bibinfo
  {journal} {J. Chem. Phys.}\ }\textbf {\bibinfo {volume} {135}},\ \bibinfo
  {pages} {124108} (\bibinfo {year} {2011})}\BibitemShut {NoStop}%
\bibitem [{\citenamefont {Khamoshi}\ \emph {et~al.}(2019)\citenamefont
  {Khamoshi}, \citenamefont {Henderson},\ and\ \citenamefont
  {Scuseria}}]{khamoshi_efficient_2019}%
  \BibitemOpen
  \bibfield  {author} {\bibinfo {author} {\bibfnamefont {A.}~\bibnamefont
  {Khamoshi}}, \bibinfo {author} {\bibfnamefont {T.~M.}\ \bibnamefont
  {Henderson}}, \ and\ \bibinfo {author} {\bibfnamefont {G.~E.}\ \bibnamefont
  {Scuseria}},\ }\href {\doibase 10.1063/1.5127850} {\bibfield  {journal}
  {\bibinfo  {journal} {J. Chem. Phys.}\ }\textbf {\bibinfo {volume} {151}},\
  \bibinfo {pages} {184103} (\bibinfo {year} {2019})}\BibitemShut {NoStop}%
\bibitem [{\citenamefont {Khamoshi}\ \emph {et~al.}()\citenamefont {Khamoshi},
  \citenamefont {Henderson},\ and\ \citenamefont
  {Scuseria}}]{khamoshi_manuscript_nodate}%
  \BibitemOpen
  \bibfield  {author} {\bibinfo {author} {\bibfnamefont {A.}~\bibnamefont
  {Khamoshi}}, \bibinfo {author} {\bibfnamefont {T.~M.}\ \bibnamefont
  {Henderson}}, \ and\ \bibinfo {author} {\bibfnamefont {G.~E.}\ \bibnamefont
  {Scuseria}},\ }\href@noop {} {\bibinfo  {journal} {Manuscript in
  preparation}\ }\BibitemShut {NoStop}%
\bibitem [{\citenamefont {Jiang}\ \emph {et~al.}(2018)\citenamefont {Jiang},
  \citenamefont {Sung}, \citenamefont {Kechedzhi}, \citenamefont
  {Smelyanskiy},\ and\ \citenamefont {Boixo}}]{jiang_quantum_2018}%
  \BibitemOpen
\bibfield  {journal} {  }\bibfield  {author} {\bibinfo {author} {\bibfnamefont
  {Z.}~\bibnamefont {Jiang}}, \bibinfo {author} {\bibfnamefont {K.~J.}\
  \bibnamefont {Sung}}, \bibinfo {author} {\bibfnamefont {K.}~\bibnamefont
  {Kechedzhi}}, \bibinfo {author} {\bibfnamefont {V.~N.}\ \bibnamefont
  {Smelyanskiy}}, \ and\ \bibinfo {author} {\bibfnamefont {S.}~\bibnamefont
  {Boixo}},\ }\href {\doibase 10.1103/PhysRevApplied.9.044036} {\bibfield
  {journal} {\bibinfo  {journal} {Phys. Rev. Applied}\ }\textbf {\bibinfo
  {volume} {9}},\ \bibinfo {pages} {044036} (\bibinfo {year} {2018})},\
  \bibinfo {note} {publisher: American Physical Society}\BibitemShut {NoStop}%
\bibitem [{\citenamefont {Coleman}(1965)}]{coleman_structure_1965}%
  \BibitemOpen
  \bibfield  {author} {\bibinfo {author} {\bibfnamefont {A.~J.}\ \bibnamefont
  {Coleman}},\ }\href {\doibase 10.1063/1.1704794} {\bibfield  {journal}
  {\bibinfo  {journal} {Journal of Mathematical Physics}\ }\textbf {\bibinfo
  {volume} {6}},\ \bibinfo {pages} {1425} (\bibinfo {year} {1965})}\BibitemShut
  {NoStop}%
\bibitem [{\citenamefont {Jordan}\ and\ \citenamefont
  {Wigner}(1928)}]{jordan_uber_1928}%
  \BibitemOpen
  \bibfield  {author} {\bibinfo {author} {\bibfnamefont {P.}~\bibnamefont
  {Jordan}}\ and\ \bibinfo {author} {\bibfnamefont {E.}~\bibnamefont
  {Wigner}},\ }\href {\doibase 10.1007/BF01331938} {\bibfield  {journal}
  {\bibinfo  {journal} {Z. Physik}\ }\textbf {\bibinfo {volume} {47}},\
  \bibinfo {pages} {631} (\bibinfo {year} {1928})}\BibitemShut {NoStop}%
\bibitem [{\citenamefont {Bravyi}\ and\ \citenamefont
  {Kitaev}(2002)}]{bravyi_fermionic_2002}%
  \BibitemOpen
  \bibfield  {author} {\bibinfo {author} {\bibfnamefont {S.~B.}\ \bibnamefont
  {Bravyi}}\ and\ \bibinfo {author} {\bibfnamefont {A.~Y.}\ \bibnamefont
  {Kitaev}},\ }\href {\doibase 10.1006/aphy.2002.6254} {\bibfield  {journal}
  {\bibinfo  {journal} {Annals of Physics}\ }\textbf {\bibinfo {volume}
  {298}},\ \bibinfo {pages} {210} (\bibinfo {year} {2002})}\BibitemShut
  {NoStop}%
\bibitem [{\citenamefont {Seeley}\ \emph {et~al.}(2012)\citenamefont {Seeley},
  \citenamefont {Richard},\ and\ \citenamefont
  {Love}}]{seeley_bravyi-kitaev_2012}%
  \BibitemOpen
  \bibfield  {author} {\bibinfo {author} {\bibfnamefont {J.~T.}\ \bibnamefont
  {Seeley}}, \bibinfo {author} {\bibfnamefont {M.~J.}\ \bibnamefont {Richard}},
  \ and\ \bibinfo {author} {\bibfnamefont {P.~J.}\ \bibnamefont {Love}},\
  }\href {\doibase 10.1063/1.4768229} {\bibfield  {journal} {\bibinfo
  {journal} {J. Chem. Phys.}\ }\textbf {\bibinfo {volume} {137}},\ \bibinfo
  {pages} {224109} (\bibinfo {year} {2012})}\BibitemShut {NoStop}%
\bibitem [{\citenamefont {Hua}(1944)}]{hua_theory_1944}%
  \BibitemOpen
  \bibfield  {author} {\bibinfo {author} {\bibfnamefont {L.-K.}\ \bibnamefont
  {Hua}},\ }\href {\doibase 10.2307/2371765} {\bibfield  {journal} {\bibinfo
  {journal} {American Journal of Mathematics}\ }\textbf {\bibinfo {volume}
  {66}},\ \bibinfo {pages} {531} (\bibinfo {year} {1944})}\BibitemShut
  {NoStop}%
\bibitem [{\citenamefont {Richardson}\ and\ \citenamefont
  {Sherman}(1964)}]{richardson_exact_1964}%
  \BibitemOpen
  \bibfield  {author} {\bibinfo {author} {\bibfnamefont {R.~W.}\ \bibnamefont
  {Richardson}}\ and\ \bibinfo {author} {\bibfnamefont {N.}~\bibnamefont
  {Sherman}},\ }\href {\doibase 10.1016/0029-5582(64)90687-X} {\bibfield
  {journal} {\bibinfo  {journal} {Nuclear Physics}\ }\textbf {\bibinfo {volume}
  {52}},\ \bibinfo {pages} {221} (\bibinfo {year} {1964})}\BibitemShut
  {NoStop}%
\bibitem [{\citenamefont {Ortiz}\ \emph {et~al.}(2001)\citenamefont {Ortiz},
  \citenamefont {Gubernatis}, \citenamefont {Knill},\ and\ \citenamefont
  {Laflamme}}]{ortiz_quantum_2001}%
  \BibitemOpen
  \bibfield  {author} {\bibinfo {author} {\bibfnamefont {G.}~\bibnamefont
  {Ortiz}}, \bibinfo {author} {\bibfnamefont {J.~E.}\ \bibnamefont
  {Gubernatis}}, \bibinfo {author} {\bibfnamefont {E.}~\bibnamefont {Knill}}, \
  and\ \bibinfo {author} {\bibfnamefont {R.}~\bibnamefont {Laflamme}},\ }\href
  {\doibase 10.1103/PhysRevA.64.022319} {\bibfield  {journal} {\bibinfo
  {journal} {Phys. Rev. A}\ }\textbf {\bibinfo {volume} {64}},\ \bibinfo
  {pages} {022319} (\bibinfo {year} {2001})},\ \bibinfo {note} {publisher:
  American Physical Society}\BibitemShut {NoStop}%
\bibitem [{\citenamefont {Elfving}\ \emph {et~al.}(2020)\citenamefont
  {Elfving}, \citenamefont {Gámez},\ and\ \citenamefont
  {Gogolin}}]{elfving_simulating_2020}%
  \BibitemOpen
  \bibfield  {author} {\bibinfo {author} {\bibfnamefont {V.~E.}\ \bibnamefont
  {Elfving}}, \bibinfo {author} {\bibfnamefont {J.~A.}\ \bibnamefont {Gámez}},
  \ and\ \bibinfo {author} {\bibfnamefont {C.}~\bibnamefont {Gogolin}},\ }\href
  {http://arxiv.org/abs/2002.00035} {\bibfield  {journal} {\bibinfo  {journal}
  {arXiv:2002.00035 [quant-ph]}\ } (\bibinfo {year} {2020})},\ \bibinfo {note}
  {arXiv: 2002.00035}\BibitemShut {NoStop}%
\bibitem [{\citenamefont {Barenco}\ \emph {et~al.}(1995)\citenamefont
  {Barenco}, \citenamefont {Bennett}, \citenamefont {Cleve}, \citenamefont
  {DiVincenzo}, \citenamefont {Margolus}, \citenamefont {Shor}, \citenamefont
  {Sleator}, \citenamefont {Smolin},\ and\ \citenamefont
  {Weinfurter}}]{barenco_elementary_1995}%
  \BibitemOpen
  \bibfield  {author} {\bibinfo {author} {\bibfnamefont {A.}~\bibnamefont
  {Barenco}}, \bibinfo {author} {\bibfnamefont {C.~H.}\ \bibnamefont
  {Bennett}}, \bibinfo {author} {\bibfnamefont {R.}~\bibnamefont {Cleve}},
  \bibinfo {author} {\bibfnamefont {D.~P.}\ \bibnamefont {DiVincenzo}},
  \bibinfo {author} {\bibfnamefont {N.}~\bibnamefont {Margolus}}, \bibinfo
  {author} {\bibfnamefont {P.}~\bibnamefont {Shor}}, \bibinfo {author}
  {\bibfnamefont {T.}~\bibnamefont {Sleator}}, \bibinfo {author} {\bibfnamefont
  {J.~A.}\ \bibnamefont {Smolin}}, \ and\ \bibinfo {author} {\bibfnamefont
  {H.}~\bibnamefont {Weinfurter}},\ }\href {\doibase 10.1103/PhysRevA.52.3457}
  {\bibfield  {journal} {\bibinfo  {journal} {Phys. Rev. A}\ }\textbf {\bibinfo
  {volume} {52}},\ \bibinfo {pages} {3457} (\bibinfo {year}
  {1995})}\BibitemShut {NoStop}%
\bibitem [{\citenamefont {Izmaylov}(2019)}]{izmaylov_construction_2019}%
  \BibitemOpen
  \bibfield  {author} {\bibinfo {author} {\bibfnamefont {A.~F.}\ \bibnamefont
  {Izmaylov}},\ }\href {\doibase 10.1021/acs.jpca.9b01103} {\bibfield
  {journal} {\bibinfo  {journal} {J. Phys. Chem. A}\ }\textbf {\bibinfo
  {volume} {123}},\ \bibinfo {pages} {3429} (\bibinfo {year}
  {2019})}\BibitemShut {NoStop}%
\bibitem [{\citenamefont {Smeyers}\ and\ \citenamefont
  {Doreste‐Suarez}(1973)}]{smeyers_half-projected_1973}%
  \BibitemOpen
  \bibfield  {author} {\bibinfo {author} {\bibfnamefont {Y.~G.}\ \bibnamefont
  {Smeyers}}\ and\ \bibinfo {author} {\bibfnamefont {L.}~\bibnamefont
  {Doreste‐Suarez}},\ }\href {\doibase 10.1002/qua.560070406} {\bibfield
  {journal} {\bibinfo  {journal} {International Journal of Quantum Chemistry}\
  }\textbf {\bibinfo {volume} {7}},\ \bibinfo {pages} {687} (\bibinfo {year}
  {1973})}\BibitemShut {NoStop}%
\bibitem [{\citenamefont {Yen}\ \emph {et~al.}(2019)\citenamefont {Yen},
  \citenamefont {Lang},\ and\ \citenamefont {Izmaylov}}]{yen_exact_2019}%
  \BibitemOpen
  \bibfield  {author} {\bibinfo {author} {\bibfnamefont {T.-C.}\ \bibnamefont
  {Yen}}, \bibinfo {author} {\bibfnamefont {R.~A.}\ \bibnamefont {Lang}}, \
  and\ \bibinfo {author} {\bibfnamefont {A.~F.}\ \bibnamefont {Izmaylov}},\
  }\href {\doibase 10.1063/1.5110682} {\bibfield  {journal} {\bibinfo
  {journal} {J. Chem. Phys.}\ }\textbf {\bibinfo {volume} {151}},\ \bibinfo
  {pages} {164111} (\bibinfo {year} {2019})}\BibitemShut {NoStop}%
\bibitem [{\citenamefont {Mihalka}\ \emph {et~al.}(2020)\citenamefont
  {Mihalka}, \citenamefont {Surjan},\ and\ \citenamefont
  {Szabados}}]{mihalka_half-projection_2020}%
  \BibitemOpen
  \bibfield  {author} {\bibinfo {author} {\bibfnamefont {Z.~E.}\ \bibnamefont
  {Mihalka}}, \bibinfo {author} {\bibfnamefont {P.~R.}\ \bibnamefont {Surjan}},
  \ and\ \bibinfo {author} {\bibfnamefont {A.}~\bibnamefont {Szabados}},\
  }\href {\doibase 10.1021/acs.jctc.9b00858} {\bibfield  {journal} {\bibinfo
  {journal} {J. Chem. Theory Comput.}\ }\textbf {\bibinfo {volume} {16}},\
  \bibinfo {pages} {892} (\bibinfo {year} {2020})}\BibitemShut {NoStop}%
\bibitem [{\citenamefont {Childs}\ and\ \citenamefont
  {Wiebe}(2012)}]{childs_hamiltonian_2012}%
  \BibitemOpen
  \bibfield  {author} {\bibinfo {author} {\bibfnamefont {A.~M.}\ \bibnamefont
  {Childs}}\ and\ \bibinfo {author} {\bibfnamefont {N.}~\bibnamefont {Wiebe}},\
  }\href {\doibase 10.26421/QIC12.11-12} {\bibfield  {journal} {\bibinfo
  {journal} {QIC}\ }\textbf {\bibinfo {volume} {12}} (\bibinfo {year} {2012}),\
  10.26421/QIC12.11-12},\ \bibinfo {note} {arXiv: 1202.5822}\BibitemShut
  {NoStop}%
\bibitem [{\citenamefont {Aharonov}\ \emph {et~al.}(2009)\citenamefont
  {Aharonov}, \citenamefont {Jones},\ and\ \citenamefont
  {Landau}}]{aharonov_polynomial_2009}%
  \BibitemOpen
  \bibfield  {author} {\bibinfo {author} {\bibfnamefont {D.}~\bibnamefont
  {Aharonov}}, \bibinfo {author} {\bibfnamefont {V.}~\bibnamefont {Jones}}, \
  and\ \bibinfo {author} {\bibfnamefont {Z.}~\bibnamefont {Landau}},\ }\href
  {\doibase 10.1007/s00453-008-9168-0} {\bibfield  {journal} {\bibinfo
  {journal} {Algorithmica}\ }\textbf {\bibinfo {volume} {55}},\ \bibinfo
  {pages} {395} (\bibinfo {year} {2009})}\BibitemShut {NoStop}%
\bibitem [{\citenamefont {Stair}\ \emph {et~al.}(2020)\citenamefont {Stair},
  \citenamefont {Huang},\ and\ \citenamefont
  {Evangelista}}]{stair_multireference_2020}%
  \BibitemOpen
  \bibfield  {author} {\bibinfo {author} {\bibfnamefont {N.~H.}\ \bibnamefont
  {Stair}}, \bibinfo {author} {\bibfnamefont {R.}~\bibnamefont {Huang}}, \ and\
  \bibinfo {author} {\bibfnamefont {F.~A.}\ \bibnamefont {Evangelista}},\
  }\href {\doibase 10.1021/acs.jctc.9b01125} {\bibfield  {journal} {\bibinfo
  {journal} {J. Chem. Theory Comput.}\ }\textbf {\bibinfo {volume} {16}},\
  \bibinfo {pages} {2236} (\bibinfo {year} {2020})}\BibitemShut {NoStop}%
\bibitem [{\citenamefont {Matsuzawa}\ and\ \citenamefont
  {Kurashige}(2020)}]{matsuzawa_jastrow-type_2020}%
  \BibitemOpen
  \bibfield  {author} {\bibinfo {author} {\bibfnamefont {Y.}~\bibnamefont
  {Matsuzawa}}\ and\ \bibinfo {author} {\bibfnamefont {Y.}~\bibnamefont
  {Kurashige}},\ }\href {\doibase 10.1021/acs.jctc.9b00963} {\bibfield
  {journal} {\bibinfo  {journal} {J. Chem. Theory Comput.}\ }\textbf {\bibinfo
  {volume} {16}},\ \bibinfo {pages} {944} (\bibinfo {year} {2020})}\BibitemShut
  {NoStop}%
\bibitem [{\citenamefont {Egger}\ \emph {et~al.}(2019)\citenamefont {Egger},
  \citenamefont {Ganzhorn}, \citenamefont {Salis}, \citenamefont {Fuhrer},
  \citenamefont {Mueller}, \citenamefont {Barkoutsos}, \citenamefont {Moll},
  \citenamefont {Tavernelli},\ and\ \citenamefont
  {Filipp}}]{egger_entanglement_2019}%
  \BibitemOpen
  \bibfield  {author} {\bibinfo {author} {\bibfnamefont {D.~J.}\ \bibnamefont
  {Egger}}, \bibinfo {author} {\bibfnamefont {M.}~\bibnamefont {Ganzhorn}},
  \bibinfo {author} {\bibfnamefont {G.}~\bibnamefont {Salis}}, \bibinfo
  {author} {\bibfnamefont {A.}~\bibnamefont {Fuhrer}}, \bibinfo {author}
  {\bibfnamefont {P.}~\bibnamefont {Mueller}}, \bibinfo {author} {\bibfnamefont
  {P.~K.}\ \bibnamefont {Barkoutsos}}, \bibinfo {author} {\bibfnamefont
  {N.}~\bibnamefont {Moll}}, \bibinfo {author} {\bibfnamefont {I.}~\bibnamefont
  {Tavernelli}}, \ and\ \bibinfo {author} {\bibfnamefont {S.}~\bibnamefont
  {Filipp}},\ }\href {\doibase 10.1103/PhysRevApplied.11.014017} {\bibfield
  {journal} {\bibinfo  {journal} {Phys. Rev. Applied}\ }\textbf {\bibinfo
  {volume} {11}},\ \bibinfo {pages} {014017} (\bibinfo {year} {2019})},\
  \bibinfo {note} {arXiv: 1804.04900}\BibitemShut {NoStop}%
\bibitem [{\citenamefont {Evangelista}\ \emph {et~al.}(2019)\citenamefont
  {Evangelista}, \citenamefont {Chan},\ and\ \citenamefont
  {Scuseria}}]{evangelista_exact_2019}%
  \BibitemOpen
  \bibfield  {author} {\bibinfo {author} {\bibfnamefont {F.~A.}\ \bibnamefont
  {Evangelista}}, \bibinfo {author} {\bibfnamefont {G.~K.-L.}\ \bibnamefont
  {Chan}}, \ and\ \bibinfo {author} {\bibfnamefont {G.~E.}\ \bibnamefont
  {Scuseria}},\ }\href {\doibase 10.1063/1.5133059@jcp.2020.EDCH2019.issue-1}
  {\bibfield  {journal} {\bibinfo  {journal} {J. Chem. Phys.}\ }\textbf
  {\bibinfo {volume} {EDCH2019}},\ \bibinfo {pages} {244112} (\bibinfo {year}
  {2019})}\BibitemShut {NoStop}%
\bibitem [{\citenamefont {Izmaylov}\ \emph {et~al.}(2020)\citenamefont
  {Izmaylov}, \citenamefont {Díaz-Tinoco},\ and\ \citenamefont
  {Lang}}]{izmaylov_order_2020}%
  \BibitemOpen
  \bibfield  {author} {\bibinfo {author} {\bibfnamefont {A.~F.}\ \bibnamefont
  {Izmaylov}}, \bibinfo {author} {\bibfnamefont {M.}~\bibnamefont
  {Díaz-Tinoco}}, \ and\ \bibinfo {author} {\bibfnamefont {R.~A.}\
  \bibnamefont {Lang}},\ }\href {\doibase 10.1039/D0CP01707H} {\bibfield
  {journal} {\bibinfo  {journal} {Phys. Chem. Chem. Phys.}\ }\textbf {\bibinfo
  {volume} {22}},\ \bibinfo {pages} {12980} (\bibinfo {year}
  {2020})}\BibitemShut {NoStop}%
\bibitem [{\citenamefont {Tannu}\ and\ \citenamefont
  {Qureshi}(2019)}]{tannu_not_2019}%
  \BibitemOpen
  \bibfield  {author} {\bibinfo {author} {\bibfnamefont {S.~S.}\ \bibnamefont
  {Tannu}}\ and\ \bibinfo {author} {\bibfnamefont {M.~K.}\ \bibnamefont
  {Qureshi}},\ }in\ \href {\doibase 10.1145/3297858.3304007} {\emph {\bibinfo
  {booktitle} {Proceedings of the {Twenty}-{Fourth} {International}
  {Conference} on {Architectural} {Support} for {Programming} {Languages} and
  {Operating} {Systems}}}},\ \bibinfo {series and number} {{ASPLOS} '19}\
  (\bibinfo  {publisher} {Association for Computing Machinery},\ \bibinfo
  {address} {Providence, RI, USA},\ \bibinfo {year} {2019})\ pp.\ \bibinfo
  {pages} {987--999}\BibitemShut {NoStop}%
\bibitem [{noa(2019)}]{noauthor_qiskit_2019}%
  \BibitemOpen
  \href {\doibase 10.5281/zenodo.2562111} {\enquote {\bibinfo {title} {Qiskit:
  {An} {Open}-source {Framework} for {Quantum} {Computing}},}\ } (\bibinfo
  {year} {2019})\BibitemShut {NoStop}%
\bibitem [{\citenamefont {Kraft}(1988)}]{kraft_software_1988}%
  \BibitemOpen
  \bibfield  {author} {\bibinfo {author} {\bibfnamefont {D.}~\bibnamefont
  {Kraft}},\ }\href {http://hdl.handle.net/10068/147127} {\emph {\bibinfo
  {title} {A software package for sequential quadratic programming}}},\
  \bibinfo {type} {Tech. Rep.}\ \bibinfo {number} {DFVLR-FB--88-28}\ (\bibinfo
  {institution} {DLR German Aerospace Center – Institute for Flight
  Mechanics},\ \bibinfo {address} {Koln, Germany},\ \bibinfo {year}
  {1988})\BibitemShut {NoStop}%
\bibitem [{\citenamefont {Byrd}\ \emph {et~al.}(1995)\citenamefont {Byrd},
  \citenamefont {Lu}, \citenamefont {Nocedal},\ and\ \citenamefont
  {Zhu}}]{byrd_limited_1995}%
  \BibitemOpen
  \bibfield  {author} {\bibinfo {author} {\bibfnamefont {R.~H.}\ \bibnamefont
  {Byrd}}, \bibinfo {author} {\bibfnamefont {P.}~\bibnamefont {Lu}}, \bibinfo
  {author} {\bibfnamefont {J.}~\bibnamefont {Nocedal}}, \ and\ \bibinfo
  {author} {\bibfnamefont {C.}~\bibnamefont {Zhu}},\ }\href {\doibase
  10.1137/0916069} {\bibfield  {journal} {\bibinfo  {journal} {SIAM J. Sci.
  Comput.}\ }\textbf {\bibinfo {volume} {16}},\ \bibinfo {pages} {1190}
  (\bibinfo {year} {1995})}\BibitemShut {NoStop}%
\end{thebibliography}%

\end{document}